\documentclass[[11pt]{article}
\usepackage{helvet,times,authblk}
\usepackage{bm,textcomp}
\usepackage{amsmath,amssymb,lastpage,bm,textcomp}
\usepackage{multicol}
\usepackage{bm}
\usepackage{graphicx}
\usepackage[round,numbers,sort&compress]{natbib}
\usepackage[margin=1in]{geometry}
\usepackage{color}

\newcommand{\con}[1]{[{\mathrm{#1}}]} 
\newcommand{\revise}[1]{\textcolor{black}{{#1}}}  

\begin{document}


\title{Shear-Induced Nitric Oxide Production by Endothelial Cells}

\author{K. Sriram$^{*}$, J. G. Laughlin$^{*}$,  P. Rangamani and D. M. Tartakovsky}
\affil{Department of Mechanical and Aerospace Engineering, University of California San Diego, 9500 Gilman Drive, La Jolla, CA 92093, USA}

\maketitle
\begin{abstract}
{We present a biochemical model of the wall shear stress (WSS)-induced activation of endothelial nitric oxide synthase (eNOS) in an endothelial cell (EC). The model includes three key mechanotransducers: mechanosensing ion channels, integrins and G-protein-coupled receptors. The reaction cascade consists of two interconnected parts. The first is rapid activation of calcium, which results in  formation of calcium-calmodulin complexes, followed by recruitment of eNOS from caveolae. The second is phosphoryaltion of eNOS by protein kinases PKC and AKT. The model also includes a negative feedback loop due to  inhibition of calcium influx into the cell by cyclic guanosine monophosphate (cGMP). In this feedback, increased nitric oxide (NO) levels cause an increase in cGMP levels, so that cGMP inhibition of calcium influx can limit NO production. The model was used to predict the dynamics of NO production by an EC subjected to a step increase of WSS from zero to a finite physiologically relevant value. Among several experimentally observed features, the model predicts a highly nonlinear, biphasic transient behavior of eNOS activation and NO production: a rapid initial activation due to the very rapid influx of calcium into the cytosol (occurring within 1 to 5 minutes) is followed by a sustained period of activation due to protein kinases.}
{Submitted }
{$^*$These two authors contributed equally to this work\\
Correspondence: dmt@ucsd.edu\\
}
\end{abstract}

\noindent
\textbf{Acronyms:} AKT, protein kinase B; $\con{Ca^{2+}}_\mathrm{c}$, $\con{Ca^{2+}}_\mathrm{s}$, $\con{Ca^{2+}}_\mathrm{e}$ and $\con{Ca^{2+}}_\mathrm{b}$, cytosolic, stored, external and buffer concentrations of calcium ions, respectively; Ca$_3$-CaM and Ca$_4$-CaM, calcium-calmodulin complexes with 3 and 4 calcium ions bound to CaM, respectively; CaM, calmodulin; CCE, capacitative calcium entry; cGMP, cyclic guanosine monophosphate; EC, endothelial cell; ECM, extracellular matrix;  eNOS, endothelial nitric oxide synthase; eNOS$_\mathrm{cav}$, eNOS bound to caveolin; eNOS$^*$, eNOS-CaM complex phosphorylated at Ser-1197; eNOS$_0$, caveolin-bound eNOS phosphorylated at Thr-495; ER, endoplasmic reticulum; FAK, focal adhesion kinase; G, active G proteins; G$_\mathrm{t}$, total G proteins; GPCR, G-protein-coupled receptors; Hsp90, heat shock protein 90; GTP, guanosine triphosphate; IP$_3$, inositol triphosphate; L-Arg, L-form of arginine; MSIC, mechanosensing ion channel; NO, nitric oxide; O$_2$, oxygen; PIP$_2$, phosphatidylinositol 4,5-bisphosphate; PIP$_3$, phosphatidylinositol (3,4,5)-triphosphate; PI3K, phosphatidylinositide 3-kinases; PKC, protein kinase C; RBC, red blood cell; sGC, soluble guanylate cyclase; WSS, wall shear stress


\section{Introduction}

Nitric oxide (NO) plays a crucial biological role in the vasculature by stimulating the relaxation of vascular smooth muscle~\cite{ignarro} and, therefore, regulating vascular resistance and blood pressure. It also has various secondary roles in the vasculature, e.g., the elimination of free radicals~\cite{wink}, and the prevention of plaque buildup~\cite{davies}. NO is produced by endothelial cells (ECs) upon exposure to mechanical forces, such as hemodynamic shear stress and intraluminal pressure~\cite{Balligand}. The mechanical stimulation of ECs triggers a complex cascade of biochemical reactions, which involves multiple cellular mechanosensors and enzymes. The ultimate target of this cascade is the activation of the endothelial nitric oxide synthase (eNOS) enzyme, which catalyzes the oxidation of the L-form of the $\alpha$-amino acid arginine (L-Arg)~\cite{Balligand, ignarro, rafikov}, resulting in the production of NO. 

Several mechanosensors are responsible for the initiation of the signal transduction cascade in an EC following mechanical stimulation~\cite{martinac, chach, ando, Shyy, storch, clark, johnsonmechanotransduction}, three of which are experimentally studied and incorporated into our model. First, the opening of mechanosensing ion channels (MSICs) mediates the influx of calcium from extracellular fluid into the cytosol~\cite{martinac}. Second, the deformation of an EC activates G-protein-coupled receptors (GPCRs), which leads to the release of calcium ions inside the EC~\cite{lemon,storch,chach}. Third, shear stress applied to an EC stimulates integrins---transmembrane receptors anchoring an EC to the extracellular matrix---which activates phosphatidylinositide 3-kinases (PI3K) enzymes~\cite{katsumi, go, morello}, leading to the release of calcium inside the EC. Our model does not account for other potential mechanosensors (e.g., sodium and potassium ion channels, lipid rafts and vesicles, cytoskeletal remodeling, signaling via cadherins and other transmembrane proteins) because experimental evidence of their importance and functionality is relatively scarce.

The majority of previous modeling studies have focused on the influx of calcium ions into ECs exposed to external shear stress~\cite{plank, plank07, comerford, wiesner, wiesner1, david}. Consequently, these and other similar models treat MSICs as the sole mechanosensor responsible for shear-induced endothelial production of NO. On the biochemical side, these models have ignored the details of eNOS phosphorylation and activation of protein kinases responsible for this phosphorylation. A recent computational model~\cite{koo} ameliorates these deficiencies by incorporating a more detailed description of the kinetics of eNOS activation by calcium-calmodulin complexes and AKT (protein kinase B), following exposure to shear stress. This model also incorporates two additional mechanosensors, G-protein coupled receptors (GPCRs) and integrins. However, it does not account for the inhibitory role of PKC (protein kinase C) or the role of cGMP in providing negative feedback to the eNOS activation cascade. In addition, an analysis for how NO production (and eNOS activation) changes for different values of shear stress was lacking. 

\revise{We build upon these and other previous efforts to develop a comprehensive model of NO production in ECs, which describes the biochemical reaction cascade induced by the three types of mechanosensors (MSICs, GPCRs, and integrins). Specifically, it relies on the experimental and modeling studies~\cite{lemon, riccobene, plank, wiesner, katsumi, go, comerford, chach} to represent both the calcium influx via MSICs and the activation of GPCRs and PI3K as functions of the applied shear stress.  Our model accounts for the formation of calcium-CaM complexes, which leads to the recruitment of eNOS into an eNOS-CaM complex; the latter is subsequently phosphorylated by AKT~\cite{Balligand}, the details of which have been ignored in previous studies~\cite{plank, plank07, comerford, wiesner, wiesner1, david}. 
It also includes a negative feedback mediated by cGMP.}

\revise{We use this model to elucidate the impact of the kinases, AKT and PKC on shear-induced NO production by ECs. The model sheds new light onto the role of calcium in the endothelium NO production, which remains an open question~\cite{Balligand, dimmelernature,kuchanfrangos}. Our analysis suggests that calcium affects both the early and later stages of NO production, albeit to different degrees.
}

\section{Mathematical Model of Endothelial NO Production}

We assume that a layer of ECs is continuously supplied with metabolic substrates sufficient for maintaining endothelial NO production. Blood flow over the endothelial layer exerts shear stress $\tau$ on the surface of the outer cell wall, which triggers a cascade of biochemical reactions resulting in endothelial NO production. The reactions are assumed to occur in the cytosol, which is treated as a well-mixed continuum, i.e., concentration gradients are ignored. In what follows, the concentration of any reactant A is denoted by [A], with units of $\mu$M.

\subsection{Activation of G proteins}

Deformation of a cell membrane by WSS activates GPCRs, which are mechanically coupled to the cell membrane and serve as force transducers that activate G proteins~\cite{storch, kuchangprotein, jogprotein}. The rate of G-protein activation is given by~\cite{lemon, riccobene}
\begin{equation}\label{gpcr}
\frac{\mathrm d \con{G} }{ \mathrm d t} = k_a [ R^* ] ( \con{G_t} - \con{G}) - k_d \con{G}, 
\end{equation}
where $\con{G}$ is the concentration of activated G-proteins; $k_a$ and $k_d$ are, respectively, the forward and backward rate constants for G-protein activation (numerical values of these and all the other reaction rate constants and model parameters are reported in Table~\ref{table:param}); $[R^*]$ is a fraction of the G-protein coupled receptors (GPCRS) activated by shear stress; and $\con{G_t}$ is the total concentration of G-proteins present in the cell. We assume the instantaneous activation of GPCRs in response to WSS $\tau$ and fit the experimental $[R^*]$ vs. $\tau $ data from Figure 4d in~\cite{chach} with a curve
\begin{equation}\label{rstar}
[ R^* ] = \tanh \left(\frac{\pi \tau}{\Lambda} \right)   
\end{equation}
where $\tau$ is in dynes/cm$^2$. The data reported in~\cite{chach} justify the assumption of instantaneous GPCR activation/inactivation: activation (or inactivation) of GPCRs in response to applied (or removed) WSS occurs on a time scale of 1~ns. 

\begin{table}[htbp]
\small
\centering
\caption{Reaction rate constants and other model parameters. An estimation procedure for the parameter values not found in the literature is discussed in section S1 of the Supplementary Material (SM).}
\label{table:kinetic_index}
\begin{tabular}{@{}lllllllllllll@{}}
\hline
Parameter & Value & Reference & Parameter & Value & Reference  \\ \hline
$k_a$ & $0.017$ s$^{-1}$ & \cite{lemon} & $k_d$ & 0.15 s$^{-1}$ & \cite{lemon} \\
$\mu_1$ & $0.2$ s$^{-1}$ & \cite{plank} &
$\alpha$ & 8.37 s$^{-1}$ & SM \\
$\Lambda$ & 15 dynes/cm$^2$ & SM &
$ K_\text{dCaM}$ & 1 $\mu$M & SM \\
$ r_r $ & 10 s$^{-1}$ & SM &
$ k_{\text{PIP}_2}^-$ & 0.7024 s$^{-1}$ & \cite{sed} & \\
$ V_r $ & $3.5$ & \cite{comerford} &
$q_\text{max}$ & 17.6 $\mu$Ms$^{-1}$ & \cite{plank} \\
$ k_\text{rel} $ & $6.64$ s$^{-1}$  & \cite{comerford} &
$ k_\text{res} $ & $5$ $\mu$Ms$^{-1}$  & \cite{comerford}  \\ 
$ k_\text{out} $ & 24.7 $\mu$Ms$^{-1}$  & \cite{comerford} &
$ k_\text{CCE} $ & $8 \times 10^{-7}$ & \cite{plank} \\
$ k_{2} $ & 0.2 $\mu$M & \cite{comerford} &
$ k_{3} $ & 0.15 $\mu$M  & \cite{comerford} \\
$ k_{5} $ & 0.32 $\mu$M & \cite{comerford} &
$ k_\text{off} $ & 300 s$^{-1} $ & \cite{plank} \\
$ k_\text{on} $ & 100 $\mu$M$^{-1}$s$^{-1}$ & \cite{plank} &
$ k_\text{leak} $ & $10^{-7}$ $\mu$M$^{-1}$s$^{-1}$ & \cite{plank} \\
$\chi$ & $28.6$ dynes/cm$^2$ & \cite{comerford, plank} &
$a_\text{PI3K}$ & $2.5$ & SM \\
$\eta$ & $0.003$ s$^{-1}$ & SM  &
$\delta$ & $24$ dynes/cm$^2$ & SM \\
$k_{1p}$ & $0.021$ s$^{-1}$ & \cite{sed} &
$k_{2p}$ & $0.022$ s$^{-1} $ & \cite{sed} \\
$k_\text{AKT}^-$ & $0.1155$ s$^{-1}$ & \cite{sed} &
$k_\text{PKC}^-$ &  $ 0.1155$ s$^{-1}$  & \cite{sed} \\
$k_\text{Thr}^+$ & 0.002 s$^{-1} $ & \cite{bredt} &
$k_\text{Thr}^-$ & $2.22 \times 10^{-4}$ s$^{-1}$ & \cite{bredt} \\
$ k_\text{Ca$_4$-CaM} $ & $100$ s$^{-1}$ & SM &
$ \beta$ & 2.7 & SM \\
$ \theta $ & $0.0045$ & SM &
$K_\text{CaM}^+$ & 7.5 s$^{-1}$ & SM \\
$K_\text{0.5CaM}$ & 3 $\mu$M & \cite{michel} &
$K_\text{CaM}^-$ & $0.01$ s$^{-1}$ & \cite{mcmurry} \\
$k_\text{eAKT}^\text{max}$ & 0.004 s$^{-1} $ & SM &
$k_\text{eAKT}^-$ & $2.22 \times 10^{-4}$ s$^{-1}$ & SM \\
$\lambda_\text{NO}$ & $382$ s$^{-1}$ & \cite{sriramno} &
$\Upsilon$ & 300 s$^{-1}$ & SM \\
$\phi $ & $9$ & \cite{takahashi}  &
$R_\text{NO}^\text{max}$ & 0.022 s$^{-1}$ & \cite{condo} \\
$b_{1}$ & 15.15 nM & \cite{condo} &
$N$ & 2 & \cite{plank} \\ 
$a_{0}$ & $1200.16$ nM$^2$ & \cite{condo} &
$a_{1}$ & $37.33$ nM & \cite{condo} \\
$g_{0}$ & $4.8$ nM$^2$ & \cite{condo} &
$g_{1}$ & $35.33$ nM & \cite{condo} \\
$X$ & $0.0695$ s$^{-1}$ & \cite{yang} &
$K_\text{cGMP}$ & $2$ $\mu$M & \cite{yang} \\
$V_\text{cGMP}^\text{max}$ & $1.26$ $\mu$Ms$^{-1}$ & \cite{yang}  &
$ \con{CaM}_\text{tot} $ & 30 $\mu$M & \cite{jafri} \\
$ \con{Ca^{2+}}_\text{e} $ & 1500 $\mu$M & \cite{plank} &
$ \con{G_\text{t}} $ & 0.33 $\mu$M & \cite{lemon,riccobene} \& SM \\
$ \con{PIP_{2_\text{t}}} $ & 10 $\mu$M & \cite{gamper} &
$ \con{eNOS}_\text{tot}$ & 0.04 $\mu$M & SM \\
$ K_{cp} $ & 0.002 $\mu$M$^{-1}$ & SM  & 
SGC0 & 0.1 $\mu$M & SM \\
$ M_\text{ATP}$ & 0.7937 & SM &
$\con{Ca^{2+}}_{s,0}$ & 2828 $\mu$M & SM \\
$\xi$ & 0.0075 $\mu$M$^{-1}$ & SM & 
$B_\text{tot}$ & 120 $\mu$M & \cite{jafri} \\
\hline 
\end{tabular}
\label{table:param}  
\end{table}

Activation of G proteins triggers the hydrolysis of phosphatidylinositol 4,5-bisphosphate (PIP$_2$) and formation of inositol triphosphate (IP$_3$) in accordance with a rate equation~\cite{lemon, plank}
\begin{equation}\label{ip3}
\frac{\mathrm d \con{IP_3} }{ \mathrm d t}=r_f  \con{PIP_2} - \mu_1 \con{IP_3}, \qquad  r_f=\alpha K_{cp} M_\text{ATP} \con{G}  
\end{equation}
where $\mu_1$ is the rate of IP$_3$ degradation, and $r_f$ is the $\con{G}$-dependent rate of IP$_3$ formation from PIP$_2$ whose parameterization with coefficients $\alpha$, $K_{cp}$ and $M_\text{ATP}$ is discussed in section S1 of the Supplementary Material. Equation~\ref{ip3} represents an IP$_3$-PIP$_2$ cycle (see~\cite{lemon} for details), in which the produced IP$_3$ degrades to an intermediate phospholipid that is then converted back to PIP$_2$. The rate of change of $\con{PIP_2}$ is described by a rate equation~\cite{lemon, sed}
\begin{equation}\label{pip2} 
\frac{\mathrm d \con{PIP_2} }{\mathrm d t} = - (r_f +r_r ) \con{PIP_2} - r_r \con{IP_3} + r_r \con{PIP_{2_t}} - k_\mathrm{PIP_2}^+ \con{PIP_2} + k_\mathrm{PIP_2}^- \con{PIP_3}
\end{equation}
in which the first three terms on the right hand side represent the cycling of PIP$_2$ to IP$_3$ and back to PIP$_2$, and the remaining two terms account for the phosphorylation of PIP$_2$ to PIP$_3$. Here $r_r$ is the rate constant of replenishment of PIP$_2$, and $k_\mathrm{PIP_2}^+$ and $k_\mathrm{PIP_2}^-$ are the forward and backward rate constants of PIP$_3$ formation from PIP$_2$.  Phosphorylation of PIP$_2$ forms PIP$_3$, a reaction which is catalyzed by activated phosphoinositide 3-kinase (PI3K) in accordance with~\cite{sed}
\begin{equation} \label{rate-eq:pip3}
\frac{\mathrm d \con{PIP_3}}{\mathrm d t} = k_\mathrm{PIP_2}^+ \con{PIP_2} - k_\mathrm{PIP_2}^- \con{PIP_3}.
\end{equation}
The catalytic role of PI3K is sensitive to the level of shear stress and is elaborated upon in section~\ref{activation}.

\subsection{Calcium signaling}

The shear stress $\tau$ causes the opening of MSICs, resulting in a sharp increase in $\con{Ca^{2+}}_\mathrm{c}$ due to the influx of calcium ions from the extracellular fluid. This and other processes affecting the calcium balance in an endothelial cell (e.g., capacitative calcium entry or CCE) are modeled by mass balance equations~\cite{plank, wiesner, comerford}
\begin{align}\label{calcium} 
\frac{\mathrm d (\con{Ca^{2+}}_\mathrm{c} + \con{Ca^{2+}}_\mathrm{b}) }{\mathrm d t} = q_\mathrm{rel} - q_\mathrm{res} - q_\mathrm{out} + q_\mathrm{in} + k_\mathrm{leak} \con{Ca^{2+}}_\mathrm{s}^2, \qquad
\frac{\mathrm d \con{Ca^{2+}}_\mathrm{s}}{\mathrm d t} = -V_r (q_\mathrm{rel} - q_\mathrm{res}  - k_\mathrm{leak} \con{Ca^{2+}}_\mathrm{s}^2 )
\end{align}
where $\con{Ca^{2+}}_\mathrm{s}$ and $\con{Ca^{2+}}_\mathrm{b}$ are the calcium concentrations in  the EC internal stores (i.e., within the smooth ER) and buffered in dissolved cytosolic proteins, respectively.
The fluxes $q_\mathrm{rel}$, $q_\mathrm{res}$ and $q_\mathrm{out}$ represent calcium release from internal stores, calcium resequestration into these internal stores and calcium efflux via the sodium-calcium exchanger~\cite{wiesner}, respectively. These concentration-dependent fluxes are given by
\begin{equation}\label{qrel}
  q_\mathrm{rel} = k_\mathrm{rel} \left ( \frac{ \con{IP_3} }{ k_2 + \con{IP_3} } \right )^{\!\!\! 3} \con{Ca^{2+}}_\mathrm{s}, \qquad
 q_\mathrm{res} = k_\mathrm{res} \left ( \frac{\con{Ca^{2+}}_\mathrm{c}}{k_3+\con{Ca^{2+}}_\mathrm{c}} \right )^{\!\!\! 2}, \qquad 
 q_\mathrm{out} = k_\mathrm{out} \frac{\con{Ca^{2+}}_\mathrm{c}}{k_5+\con{Ca^{2+}}_\mathrm{c}}
\end{equation}
where $k_\mathrm{rel}$, $k_\mathrm{res}$, $k_\mathrm{out}$, $k_1$, $k_3$ and $k_5$ are rate constants. Influx of calcium ions from the extracellular fluid into the cytosol occurs through both MSICs~\cite{comerford, plank} and CCE~\cite{putney}. The corresponding calcium fluxes are related by
\begin{equation}\label{qin}  
q_\mathrm{in} = q_\mathrm{MSIC} + q_\mathrm{CCE}.
\end{equation}
Following~\cite{wiesner, plank, comerford}, we assume that the rate of calcium influx via MSICs, $q_\mathrm{MSIC}$, is linearly proportional to the fraction $f_0$ of ion channels open at a given WSS $\tau$,
\begin{equation} 
q_\mathrm{MSIC} = f_0 (\tau ) \, q_\mathrm{max} = \frac{q_\mathrm{max}}{1 + N \, \exp(-W)}, \qquad  W(\tau) = W_0 \frac{(\tau + \sqrt{16 \chi^2 + \tau^2 } - 4\chi )^2}{\tau + \sqrt{16 \chi ^2 + \tau^2 }} \label{ } \end{equation}
where $q_\mathrm{max}$ is the maximal rate of influx, $W(\tau)$ quantifies the extent to which the applied mechanical force is converted to gating energy for MSICs, and the constant $\chi = 28.6$ dynes/cm$^2$  represents the membrane shear modulus. The flux due to CCE, $q_\mathrm{CCE}$, is caused by the depletion of internal calcium stores, which induces influx of calcium from the extracellular fluid \cite{putney}. The magnitude of $q_\mathrm{CCE}$ is affected by cGMP~\cite{Kwan, plank07}, such that \cite{plank, plank07, comerford}
\begin{equation}\label{cce}
q_\mathrm{CCE} =  k_\mathrm{CCE}  \psi ( \con{cGMP}) ( \con{Ca^{2+}}_\mathrm{s, 0} - \con{Ca^{2+}}_\mathrm{s} ) ( \con{Ca^{2+}}_\mathrm{e} - \con{Ca^{2+}}_\mathrm{c} ) 
 \end{equation}
where $\con{Ca^{2+}}_\mathrm{e}$ is the calcium concentration in extracellular fluid (assumed to remain constant), and $\con{Ca^{2+}}_\mathrm{s, 0}$ is the stored calcium concentration under basal conditions. The decreasing function $\psi(\con{cGMP})$ accounts for the inhibition of CCE observed in~\cite{Kwan}, where it is suggested that $\psi(\con{cGMP})$ is a linear function of the cGMP concentrations,
\begin{equation}\label{psi}
\psi = 1.00 -  \xi \con{cGMP},
\end{equation}
over the range of $\con{cGMP}$ relevant to our model, i.e., on the order of 10 $\mu$M typical of ECs and smooth muscle~\cite{yang, condo}. Here $\xi$ is a constant of proportionality, whose value is estimated in section S1 of the Supplementary Material and reported in Table~\ref{table:param}. In addition to inhibition of CCE, increase in $\con{cGMP}$ also results in inhibition of calcium transport via MSICs~\cite{Kwan}. However, the relevant data suggest that at $\con{cGMP}$ on the order 10 $\mu$M and below, this MSICs inhibition is negligible~\cite{Kwan}. We therefore ignore the role of cGMP inhibition on MSIC function.

Equation~\ref{calcium} includes calcium concentration in the ``buffer", $\con{Ca^{2+}}_\mathrm{b}$. As the  cytosolic-calcium concentration $\con{Ca^{2+}}_\mathrm{c}$ increases, it  forms calcium complexes with cytosolic proteins in accordance with a rate law~\cite{plank, jafri}
\begin{equation} 
\frac{\mathrm d \con{Ca^{2+}}_\mathrm{b} }{\mathrm d t} = k_\mathrm{on} \con{Ca^{2+}}_\mathrm{c} ( B_\text{tot} - \con{Ca^{2+}}_\mathrm{b} ) - k_\mathrm{off} \con{Ca^{2+}}_\mathrm{b}, 
\end{equation}
where $B_\text{tot}$ is the total concentration of calcium binding sites in cytosolic buffer proteins; and $k_\mathrm{on}$ and $k_\mathrm{off}$ are forward and backward buffering rate constants, respectively.

\subsection{Activation of protein kinases} 
\label{activation}

Integrins, which anchor an EC to the extracellular matrix (ECM), are connected to focal adhesion sites within the EC and act as mechanotransducers. Application of mechanical forces results in the tyrosine phosphorylation of focal adhesion kinases (FAKs), ultimately triggering the stimulation of PI3K. In ECs, this process has been studied \textit{in vitro}, where the activation of integrins, FAKs, and PI3K are all measured as functions of mechanical stimulation~\cite{go, katsumi, morello}. 

Upon exposure to shear stress, PI3K is phosphorylated rapidly, reaching maximal activation on the order of 10~s~\cite{li, go}. This process is caused by activation of the FAK/Src complex through \cite{guan} integrins which serve as force transducers that mediate the mechanical signal. Since this time scale is an order of magnitude smaller than that of the other chemical reactions in our model (which generally occur at time scales of order 1-\revise{100} min), we neglect the time lag between the application of mechanical force and PI3K activation; application of force thus results in immediate activation of PI3K. The active PI3K species, PI3K$^*$, then gradually returns to its basal levels of activity, $\con{PI3K^*}_\mathrm{bas}$. This process is modeled as
\begin{equation}\label{pi3k}
\frac{ \con{PI3K^*} }{ \con{PI3K^*}_\mathrm{bas}} = 1 + a_\mathrm{PI3K} \, \tanh \left( \frac{\pi \tau }{\delta } \right) \, \mathrm{e}^{ - \eta t }
\end{equation}
where the constants $a_\mathrm{PI3K}$, $\delta$ and $\eta$ were fitted to the experimental data from~\cite{go, katsumi}. The data reported in \cite{go} suggest that the decay of $\con{PI3K^*}$ to its basal level occurs on a relatively fast time scale of about 5 min, while the corresponding data in~\cite{katsumi} support a larger time scale of about 30 min. In Eq.~\ref{pi3k}, this time scale is controlled by the parameter $\eta$.  To account for the time-scale variability observed in~\cite{go, katsumi}, we consider a range of $\eta$ values. Figure~S1 in the Supplemental Material reveals that the model predictions are relatively insensitive to variations in the value of $\delta$; this suggests that the transient effects of PI3K activation on NO production are negligible. Finally, activation of PI3K increases the rate of PIP$_2$ phosphorylation to PIP$_3$ \cite{sed},
\begin{equation} \label{kpip2}
k_\mathrm{PIP2}^+ = k_{1p} \frac{ \con{PI3K^*} }{ \con{PI3K^*}_\mathrm{max}} + k_{2p} = \frac{ k_{1p} }{ 1 + a_\mathrm{PI3K} } \left[ 1 + a_\mathrm{PI3K} \, \mathrm{e}^{ - \eta t } \, \tanh \left( \frac{\pi \tau }{\delta } \right)  \right] + k_{2p},
\end{equation}
thus playing a catalytic role in Eq.~\ref{rate-eq:pip3}. The values of reaction rate constants $k_{1p}$ and $k_{2p}$ are given in~\cite{sed}, and it follows from Eq.~\ref{pi3k} that the maximum concentration of active PI3K is $ \con{PI3K^*}_\mathrm{max} = \con{PI3K^*}_\mathrm{bas} ( 1 + a_\mathrm{PI3K} )$. 

PIP$_3$ meditates phosphorylation and subsequent activation of protein kinases AKT and PKC, which in turn phosphorylate eNOS~\cite{sed}. This activation is modeled by rate laws~\cite{sed}
\begin{equation}\label{eq:15}
\frac{\mathrm d \con{AKT^*} }{ \mathrm d t} = k_\mathrm{AKT}^+ \con{AKT}- k_\mathrm{AKT}^- \con{AKT^*} , \qquad \frac{\mathrm d \con{PKC^*} }{ \mathrm d t} = k_\mathrm{PKC}^+ \con{PKC} - k_\mathrm{PKC}^- \con{PKC^*}
\end{equation}
where $\con{AKT^*}$, $\con{AKT}$, $\con{PKC^*}$ and $\con{PKC}$ are concentrations of the activated (phosphorylated) AKT, unactivated AKT, activated (phosphorylated) PKC and unactivated PKC, respectively. The total concentrations of AKT and PKC are conserved, $\con{AKT} + \con{AKT^*} = \con{AKT}_\text{tot}$ and $\con{PKC} + \con{PKC^*} = \con{PKC}_\text{tot}$. The rate constants in Eqs.~\ref{eq:15} increase with $\con{PIP_3}$ in accordance with~\cite{sed}
\begin{equation}\label{eq:rates}
k_\mathrm{AKT}^+ = 0.1 k_\mathrm{AKT}^- \frac{ \con{PIP_3} - \con{PIP_3}_\text{min} }{ \con{PIP_3}_\text{max} - \con{PIP_3}_\text{min} }, \qquad
k_\mathrm{PKC}^+ = 0.1 k_\mathrm{PKC}^- \frac{ \con{PIP_3} - \con{PIP_3}_\text{min} }{ \con{PIP_3}_\text{max} - \con{PIP_3}_\text{min} }
\end{equation}
where $k_\mathrm{AKT}^-$ and $k_\mathrm{PKC}^-$ are rate constants, $\con{PIP_3}_\text{min} = 0.0031 \con{PIP_2}_\text{tot}$, and $\con{PIP_3}_\text{max} = 0.031 \con{PIP_2}_\text{tot}$.  The significant transient behavior of $\con{PIP_3}$ (Fig.~S8 in the Supplemental Material) renders the rate constants $k_\mathrm{AKT}^+$ and $k_\mathrm{PKC}^+$ time-dependent. This necessitates a numerical solution of Eqs.~\ref{eq:15}.

\subsection{Phosphorylation and formation of eNOS complexes} 
\label{complex}

Calcium forms several complexes with calmodulin (CaM), which then recruit eNOS into an eNOS-CaM complex; the presence of heat shock protein 90 (Hsp90) enhances the recruitment rate. Of the various Ca$^{2+}$/CaM complexes only Ca$_3$CaM and Ca$_4$-CaM appear to actively recruit eNOS. Among the two, Ca$_4$-CaM is the dominant species both in terms of cytosolic concentration~\cite{park} and affinity for eNOS~\cite{mcmurry}. Hence only the role of Ca$_4$-CaM in the recruitment of eNOS is accounted for in our model. We use the Hill function to relate the concentration of free Ca$_4$-CaM in the cytosol to the cytosolic calcium concentration~\cite{pereschini, porumb}, 
\begin{equation} \label{ca4cam} 
\frac{ \mathrm d \con{\text{Ca$_4$-CaM}} }{ \mathrm d t} = k_\text{Ca$_4$-CaM} \left( \theta \frac{\con{Ca^{2+}}_\mathrm{c}^\beta}{K_\mathrm{dCaM} +\con{Ca^{2+}}_\mathrm{c}^\beta} \con{CaM}_\mathrm{tot} - \con{\text{Ca$_4$-CaM}} \right).
\end{equation}
Here $\beta$ is the Hill coefficient, $ K_\mathrm{dCaM}$ is the apparent Ca$_4$-CaM dissociation constant, $k_\text{Ca$_4$-CaM}$ is the reaction rate constant, and $\con{CaM}_\mathrm{tot}$ is the total concentration of CaM in the cytosol. The coefficient $\theta$ determines the limiting amount of free Ca$_4$-CaM in the cytosol at steady state, at large cytosolic calcium  concentrations $\con{Ca^{2+}}_\mathrm{c}$. 

We model activation of eNOS from its basal (inactive) state to its fully active (Ser-1197 phophorylated) state as a three-step process~\cite{Balligand, Shyy, dudzinski, mount, takahashi}.
\paragraph{Step 1.} Caveolin-bound eNOS (eNOS$_\mathrm{cav}$) forms a complex with Ca$_4$-CaM, which we denote by eNOS-CaM. 
The rate of its formation is modeled using Michelis-Menten kinetics,
\begin{align}\label{eq:eNOS-Cam}
\frac{ \mathrm d [\text{eNOS-CaM}] }{ \mathrm d t} = \frac{k_\text{CaM}^+ \con{Ca_4CaM} }{k_{0.5\text{CaM}}+\con{Ca_4CaM} } \con{eNOS_{cav}}
- k_\text{CaM}^- [\text{eNOS-CaM}] 
- \frac{ \mathrm d [\text{eNOS-CaM}^*] }{ \mathrm d t},
\end{align}
where the forward ($k_\text{CaM}^+$) and backward ($k_\text{CaM}^-$) rate constants, and the Michaelis-Menten constant  $k_{0.5\text{CaM}}$, are obtained from the data presented in~\cite{ju, mcmurry, michel}; and eNOS-CaM$^*$ denotes the eNOS-CaM complex stabilized by phosphorylation due to AKT. 

\paragraph{Step 2.} The latter process is assumed to follow first-order kinetics,
\begin{align}\label{eq:step2}
\frac{\mathrm d [\text{eNOS-CaM}^*] }{ \mathrm d t} = k_\text{eAKT}^+ [\text{eNOS-CaM}] - k_\text{eAKT}^-[\text{eNOS-CaM}^*]
\end{align}
where $k_\text{eAKT}^+ = k_\text{AKT}^\text{max} \con{AKT^*} / \con{AKT}_\text{tot}$. The values of rate constants $k_\text{eAKT}^-$ and $k_\text{AKT}^\text{max}$  are estimated in section S1 of the Supplementary Material from the data presented in~\cite{takahashi}. The stabilized complex eNOS-CaM$^*$ complex is significantly more active than eNOS-CaM in stimulating NO production due to the catalysis of the L-Arg oxidation reaction~\cite{Balligand, Shyy, dudzinski, mount, takahashi}. 

\paragraph{Step 3.} Phosphorylation of eNOS$_\mathrm{cav}$ by PKC (at Thr-495) inhibits this activation process by blocking the formation of eNOS-CaM~\cite{Balligand}; caveolin-bound eNOS in this inactivated state is denoted by eNOS$_\mathrm{cav}^0$. We assume that eNOS phosphorylation by PKC follows first-order kinetics, 
\begin{align}\label{eq:eNOScav}
 \frac{\mathrm d \con{eNOS_{cav}^0} }{ \mathrm d t} = k_\text{Thr}^+ \frac{\con{PKC^*}}{\con{PKC}_\text{tot}}  \con{eNOS_{cav}} - k_\text{Thr}^- \con{eNOS_{cav}^0} ,
\end{align}
where the forward ($k_\text{Thr}^+$) and backward ($k_\text{Thr}^-$) rate constants are estimated from the data in~\cite{bredt}.

The total amount of eNOS in different complexes is conserved, so that the total concentration of eNOS, $\con{eNOS}_\text{tot}$, remains constant throughout these transformations. This yields a constraint on the concentrations of eNOS in different complexes~\cite{koo},
\begin{equation} \label{eq:enosconservation}
\con{eNOS_{cav}} + \con{eNOS_{cav}^0} + [\text{eNOS-CaM}] + [\text{eNOS-CaM}^*] = \con{eNOS}_\text{tot}.
\end{equation}

Two simplifications underpin our model of eNOS activation. First, the rate of dephosphorylation is assumed constant, even though dephosphorylation is mediated by various phosphatases~\cite{Balligand, greif}. We adopted this assumption because of the paucity of data on whether and how changes in shear stress affect the activation of phosphatases, such as PP2A and calcineurin. Second, the kinetics of eNOS activation is assumed to be insensitive to fluctuations in Hsp90 activity even though formation of eNOS-CaM is mediated by Hsp90~\cite{Balligand, takahashi}. This assumption is reasonable because Hsp90 is usually present in large excesses, on the order of 100 $\mu$M~\cite{nollen}, over the concentrations of other reactants, such as eNOS and CaM.

\subsection{NO production}

Once formed, the eNOS-CaM and eNOS-CaM$^*$ complexes catalyze the oxidation of L-Arg, resulting in production of NO. The latter process is modeled with the rate law,
\begin{equation} \label{eq:noprod}
\frac{\mathrm d \con{NO} }{ \mathrm d t} = Q_\text{NO} - Q_\text{sGC} - \lambda _\text{NO} \con{NO},
\end{equation}
which represents a balance between NO production (with rate $Q_\text{NO}$) and consumption. The latter is due to both NO scavenging by soluble guanylate cyclase at rate $Q_\text{sGC}$~\cite{condo} and NO  metabolism by RBCs adjacent to the endothelial cells at rate $\lambda _\text{NO}$~\cite{sriramno}. 
Following~\cite{buerk}, we use Michaelis-Menten kinetics with constant O$_2$ supply to model NO production due to oxidation of L-Arg,
\begin{align}\label{qno}
 Q_\text{NO} = R_\text{NO} \frac{ \con{O_2} }{ k_\text{mNO} + \con{O_2} }.
\end{align}
Here $k_\text{mNO}$ is the Michaelis-Menten constant and $R_\text{NO}$ is the rate of NO production, which depends on the concentrations of both phosphorylated and unphosphorylated eNOS-CaM such that
\begin{align}\label{rno} 
R_\text{NO} = k_\text{eNO} \left( [\text{eNOS-CaM}] + \phi [\text{eNOS-CaM}^*] \right) 
\end{align}
where $\phi$ indicates the extent to which AKT phosphorylation increases eNOS activity in the presence of Hsp90. Assuming a constant supply of L-arg and other substrates such as BH4 and NADH, the rate $k_\text{eNO}$ remains constant. For a constant oxygen supply,
\begin{align}\label{o2const} 
k_\text{eNO}  \frac{ \con{O_2} }{ k_\text{mNO} + \con{O_2} } = \Upsilon = \text{constant}. 
\end{align}

The experimental evidence reported in~\cite{dimmelernature, schmidt2005endostatin, ahluwalia} suggests functional presence of sGC in endothelial cells. We adopt the model~\cite{condo}, developed for smooth muscle cells, to account for the possibility of NO being scavenged by soluble guanylate cyclase (sGC) to produce sGC-NO, which catalyzes  production of cGMP from GTP,
\begin{align} \label{eq:sgc}
 Q_\text{sGC} = R_\text{NO}^\text{max}  \frac{ \con{NO}^2 + b_1 \con{NO} }{ a_0 + a_1 \con{NO} + \con{NO}^2 } \con{sGC}
\end{align}
where $b_1$, $a_0$, $a_1$ and $R_\text{NO}^\text{max}$ are constants whose values are obtained in~\cite{condo} by fitting NO - cGMP kinetics to experimental data. sGC is activated upon consuming NO, thereby stimulating the conversion of GTP into cGMP. The rate of cGMP production is expressed as a function of NO concentration~\cite{condo},
\begin{align}\label{eq:last}
\frac{\mathrm d \con{cGMP} }{ \mathrm d t} = V_\text{cGMP}^\text{max} \frac{g_0 + g_1 \con{NO} + \con{NO}^2 }{ a_0 + a_1 \con{NO} + \con{NO}^2} - \frac{ X \con{cGMP}^2 }{ k_\text{cGMP} + \con{cGMP}}
\end{align}
where $V_\text{cGMP}^\text{max}$, $g_0$, $g_1$ $X$ and $k_\text{cGMP}$ are constants whose values are obtained in~\cite{condo} by fitting NO-cGMP kinetics to experimental data. Increase in the cGMP concentration, $\con{cGMP}$, provides a negative feedback by reducing the capacitative calcium entry (CCE) into the cell in accordance with Eq.~\ref{cce}.

\begin{figure}[htbp]
\centering
\includegraphics[width=\textwidth]{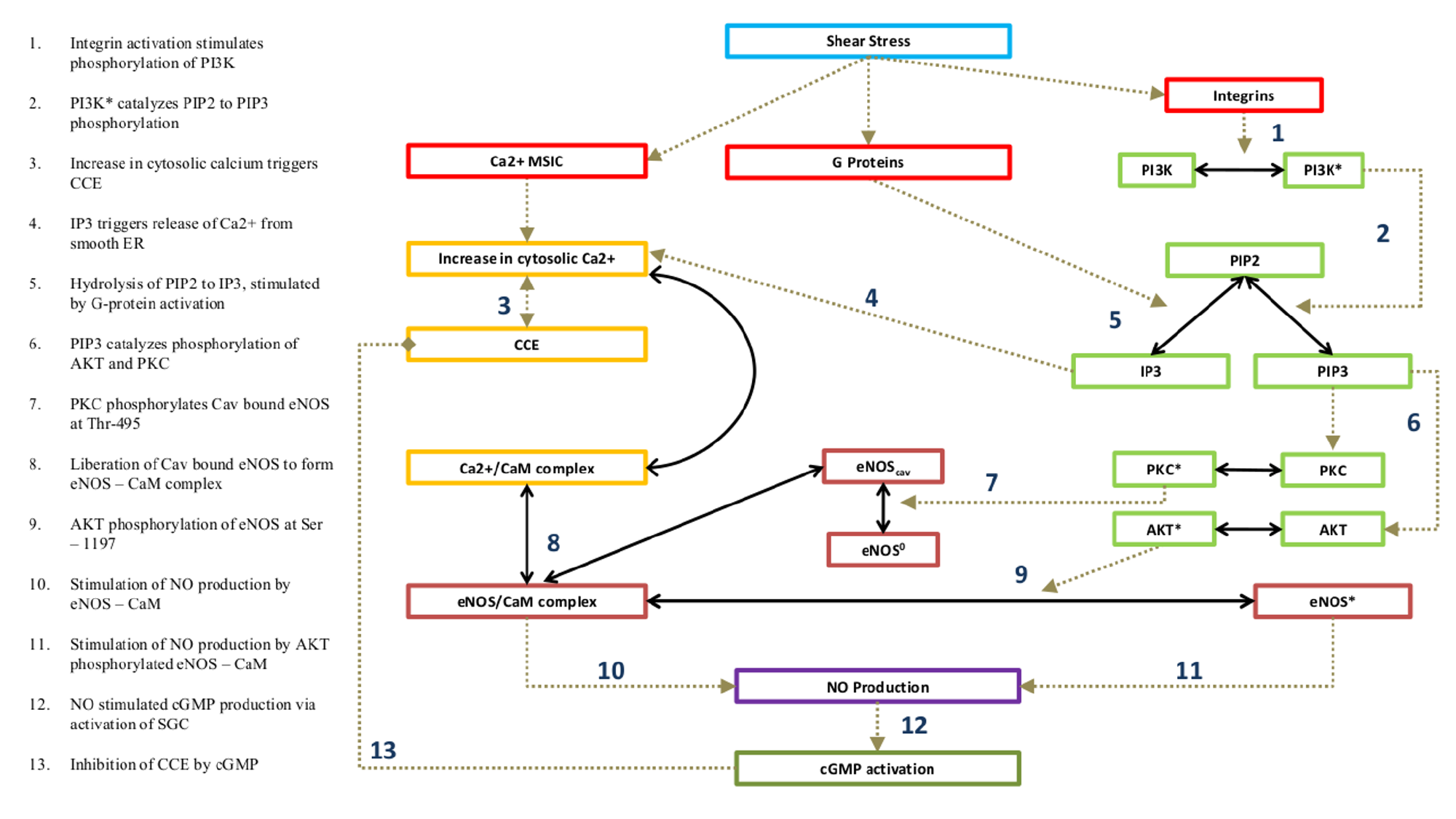}
\caption{\small Reaction network for shear-induced NO production.}\label{cascade}
\end{figure}

A system of coupled Eqs.~\ref{gpcr}--\ref{eq:last} constitutes a model of eNOS activation and NO production in response to mechanical stimulation of an endothelial cell by wall shear stress. The reaction network formed by these equations is shown in Figure~\ref{cascade}.  \revise{The numerical solution of these equations is detailed  in section S3 of the Supplementary Material.}

\section{Results}

\begin{figure}[htbp]
\centering
\includegraphics[width=\textwidth]{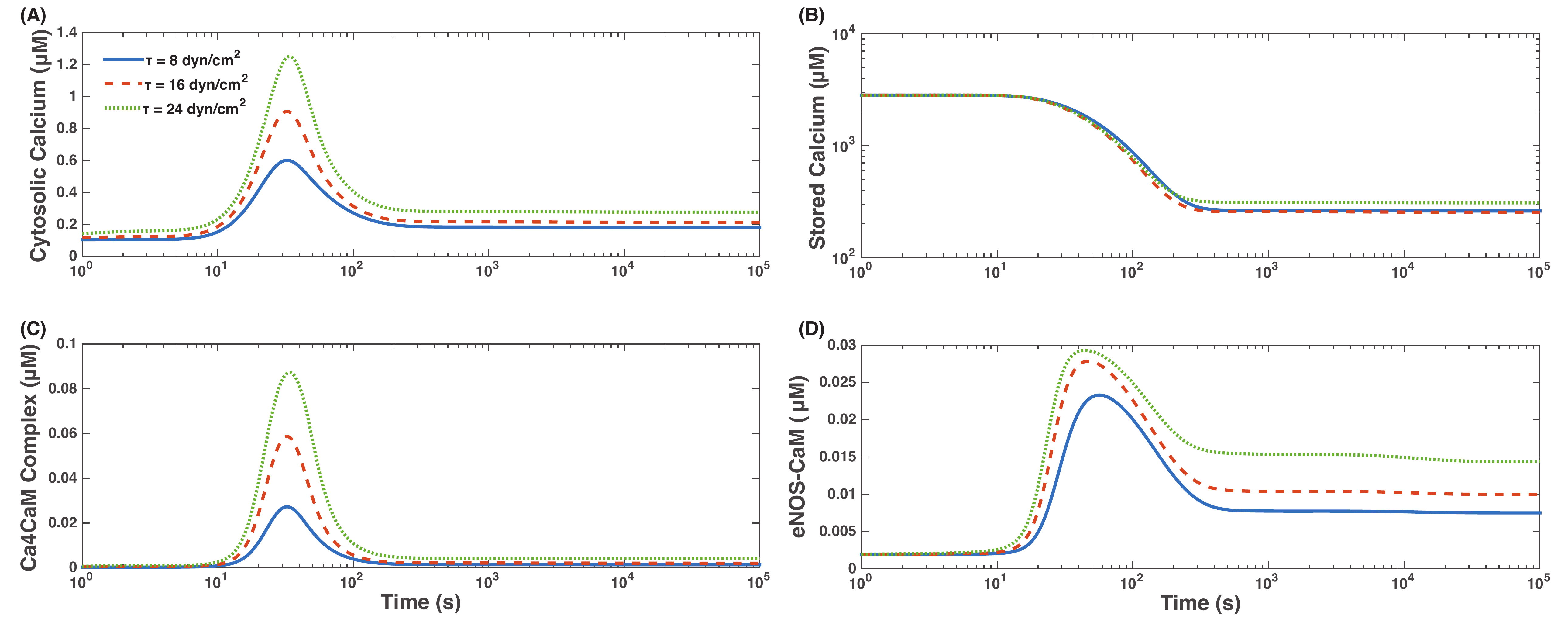}
\caption{Temporal variability of the concentrations of cytosolic calcium, $\con{Ca^{2+}}_\text{c}$, stored calcium, $\con{Ca^{2+}}_\text{s}$, and Ca$_4$CaM and eNOS-CaM complexes, $\con{Ca_4CaM}$ and $[\text{eNOS-CaM}]$, for WSS $\tau = 8$, 16 and 24 dynes/cm$^2$.}\label{calciumcam}
\end{figure}

Figure~\ref{calciumcam} shows the dynamics of concentrations of cytosolic calcium ($\con{Ca^{2+}}_\text{c}$), stored calcium ($\con{Ca^{2+}}_\text{s}$), calcium calmodulin complex ($\con{Ca_4CaM}$) and the eNOS-CaM complex ($\text{eNOS-CaM}]$) for different levels of the applied WSS. The step increase in WSS $\tau$, \revise{at time $t=0$}, induces a quick discharge of internal calcium stores within the cell as calcium enters the cytosolic volume. This results in a rapid spike in calcium levels, followed by a more gradual decline to a steady-state value of $\con{Ca^{2+}}_\text{c}$, which exceeds its basal-state counterpart. This predicted behavior is consistent with the observations~\cite{ando, comerford, helmlinger, wiesner}. The spike in $\con{Ca^{2+}}_\text{c}$ causes a corresponding increase in $\con{Ca_4CaM}$, leading to the formation of the eNOS-CaM complex. \revise{Unphosphorylated eNOS-CaM can lead to NO production, which is traditionally viewed as a calcium-independent part of the cascade. Furthermore, eNOS-CaM undergoes phosphorylation by AKT, which leads to further NO production at the longer time scale, and is also thought to be calcium-independent.  The transients for the initial calcium-dependent stage of the reaction cascade occur very fast (on the order of 100 s); the subsequent, kinase-dependent portion of the eNOS activation cascade proceeds at a more gradual pace. This kinase-dependent eNOS activation phase is also calcium-dependent through the eNOS-CaM complex.} Throughout their time course the concentrations of all forms calcium increase with WSS $\tau$ due to the role of the MSICs and G proteins.

\begin{figure}[htbp]
\centering
\includegraphics[width=\textwidth]{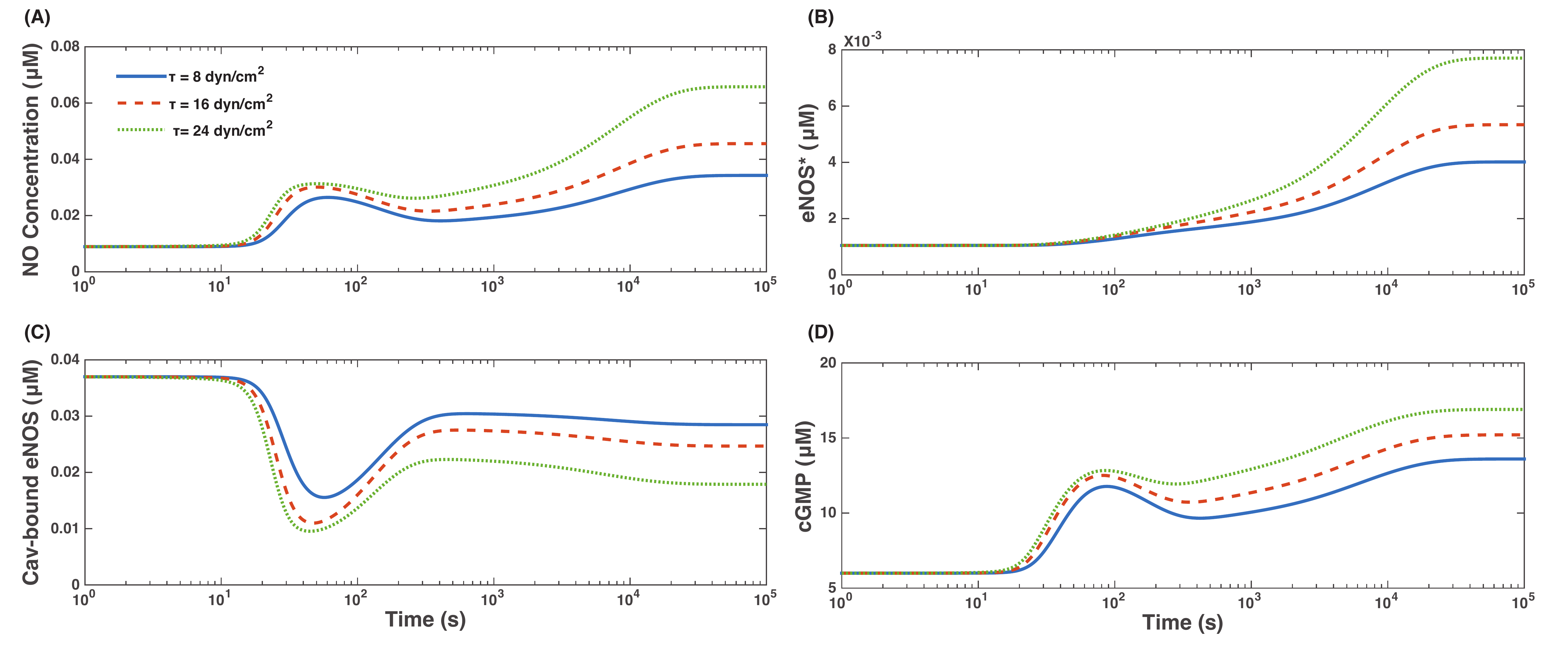}
\caption{\small Temporal variability of $\con{NO}$, AKT-phosphorylated eNOS concentration, $\con{eNOS^*}$, caveolin-bound eNOS concentration, $\con{eNOS_{cav}}$, and $\con{cGMP}$ for three levels of WSS, $\tau = 8$, 16 and 24 dynes/cm$^2$.}\label{enoscamakt}
\end{figure}

Figure~\ref{enoscamakt} shows the dynamics of AKT phosphorylation of eNOS-CaM,  forming eNOS$^*$. The increased concentrations of eNOS$^*$ form gradually, over longer time scales than the initial calcium transients. The eNOS bound to caveolin (and hence inactivated) rapidly decreases following the initial calcium influx, before reaching a steady state. The extent of depletion of the eNOS-Cav complex increases with $\tau$, as do the increased concentrations of eNOS$^*$. This shear-induced NO production, causes $\con{NO}$ to display bimodality  over time: an initial peak corresponding to activation by calcium is followed by a second, prolonged peak reflecting the role of protein kinases in eNOS activation. This predicted behavior is in agreement with the observations~\cite{Balligand, mount, Shyy} of dual nature of eNOS enzyme's activation, which is partly calcium-dependent and partly AKT-dependent. Figure~\ref{enoscamakt} also shows that $\con{cGMP}$ increases with $\con{NO}$, thereby facilitating the vasodilatory role of NO.

\subsection{Model Validation}

\begin{figure}[htbp]
\centering
\includegraphics[width=\textwidth]{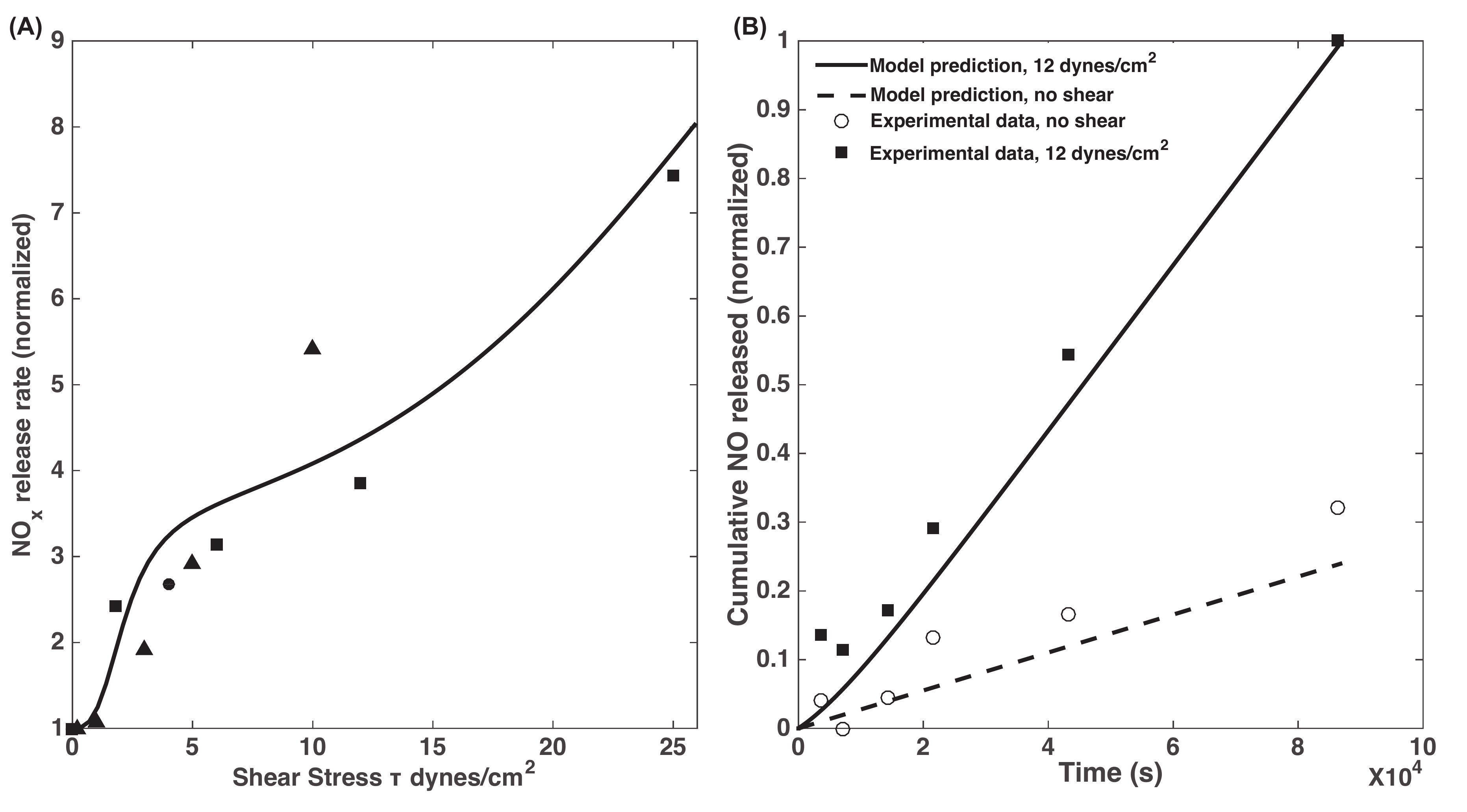}
\caption{ \revise{\textbf{(A)} Predicted (line) and observed (symbols) dependence of NO production rate on WSS. Predicted NO production rates are given by the $Q_\text{NO}$ term in Eq. \ref{eq:noprod}, which at steady state is equal to the rate of release of NO by ECs and formation of NO metabolites in the surrounding blood/media (since NO consumption by the ECs themselves was found to be negligible). Experimental data are from the following sources. Squares are from column C of Table 1 in~\cite{kuchanfrangos}, where NO production rates were estimated using measurements of NO$_x$ accumulation rates; Circles are from figure 4C in~\cite{kaur}, where NO production rates were estimated from nitrite accumulation rates; Triangles are from figure 8 in~\cite{kanai}, where NO production rates were estimated from direct measurements of moles of NO released per unit time. Each experimental data set was normalized  to the rate at $\tau = 0$, except for \cite{kanai} where the values were normalized to the lowest non-zero measurement, at $\tau = 0.2$~dynes/cm$^2$; the simulation results were normalized with the predicted rate at $\tau = 0$. \textbf{(B)} The predicted (lines) and observed (symbols) cumulative release of NO to the media/bloodstream as a function of time. The experimental data are from the top panel of figure 1 in~\cite{tsao} showing normalized increase (above baseline measurement) of NO$_x$ accumulation in conditioned media. Both experimental and model data are normalized against cumulative NO/NO$_x$ release at 12 dynes/cm$^2$ after 24 hours.}}
\label{newfigure4}
\end{figure}

\revise{To verify that our model reproduces both realistic values of NO production and realistic transient behavior, we compare the model predictions with experimental observations. Since NO is unstable, many experiments report the accumulation of nitrites and nitrates concentration as NO$_x$. We compare the NO production rates estimated experimentally (typically inferred from measurements of NO$_x$ accumulation rates) and predicted with our model (Eq. \ref{eq:noprod}, where the $Q_\text{NO}$ term gives NO production rate). Since the amount of NO metabolized by the ECs themselves ($Q_\text{sG}$) is found to be small, the NO production rate is equivalent to the amount of NO that is released and scavenged by surrounding blood/media into NO metabolites. This fact justifies the use of NO$_x$ accumulation rates as a proxy for NO production rates in  experimental studies; we, therefore, compare these values with our model simulations.The predicted and observed dependences of the rate of NO production at steady state following exposure to a range of WSS values are in general agreement with the experimental data (Fig.~\ref{newfigure4}A). Likewise, the cumulative amount of NO released by the endothelial cells (calculated using the first two terms in Eq.~\ref{eq:noprod}, which gives the amount of NO produced by the endothelial cell, minus the amount metabolized by the cell itself, and integrated over time) reproduces the observed accumulation of NO (and its metabolites) in conditioned media following application of WSS~\cite{tsao}, and when compared vs NO release/accumulation in the absence of WSS (Fig.~\ref{newfigure4}B). Despite of considerable scatter in the experimental data, our model predictions are broadly compatible with the observations.}

\begin{figure}[htbp]
\centering
\includegraphics[width=\textwidth]{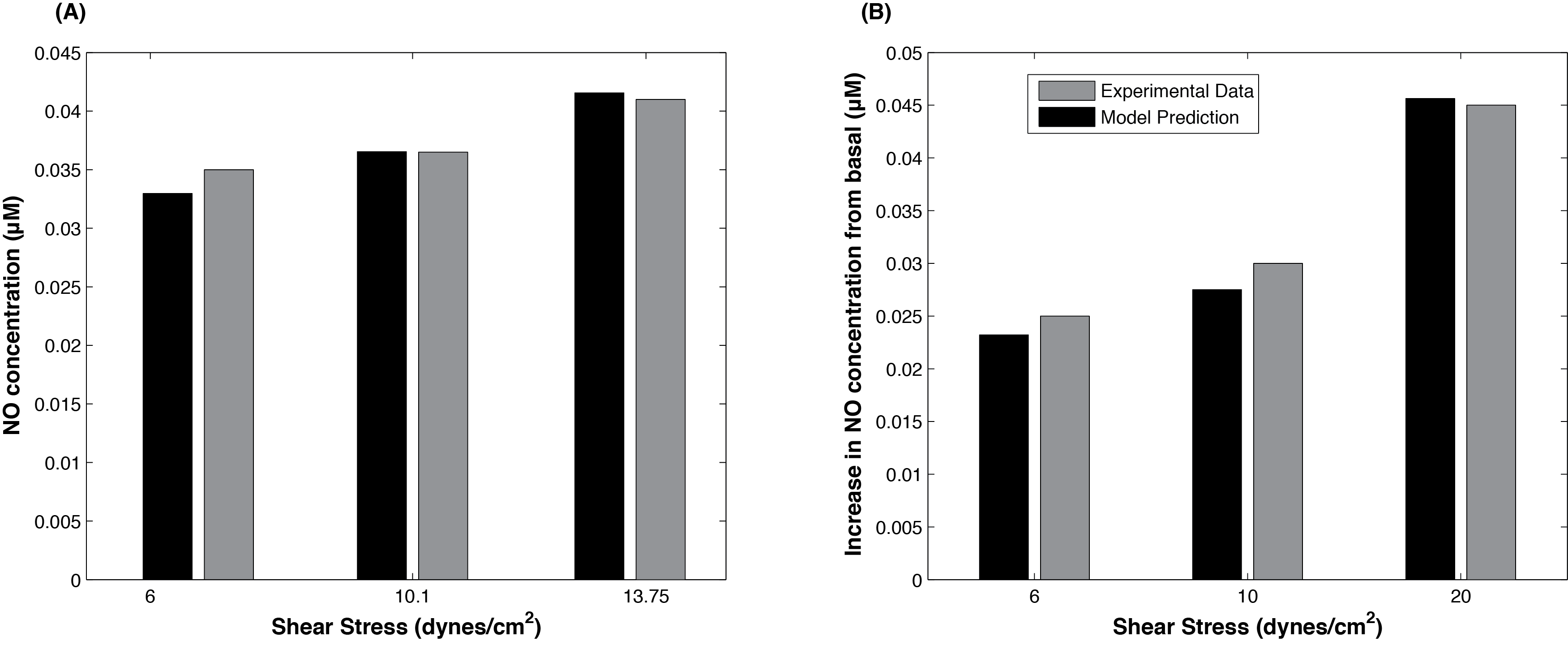}
\caption{\small (A)  The predicted and observed NO concentration at three levels of WSS $\tau$ (in dynes/cm$^2$). The experimental data are from~\cite{mashour}. (B) The predicted and observed changes in NO concentration from its basal levels for three values of WSS (in dynes/cm$^2$). The experimental data are from~\cite{andrews}.}\label{noconcmash}. 
\end{figure}

Figure \ref{noconcmash}(A, B) demonstrates the model's ability to predict steady state NO concentrations at three levels of WSS $\tau$. The predicted values of $\con{NO}$ are within 5\% of the experimental observations reported in~\cite{mashour}. This figure provides a quantitative justification for the  assumption of a linear relationship between $\con{NO}$ and $\tau$, which is routinely used in microcirculation models (see, e.g.,~\cite{sriramautoreg} and the references therein). While this relationship is clearly nonlinear, it can be treated as linear over the physiologically relevant range of WSS shown in Figure~\ref{noconcmash}. As a final exercise in model validation, Figure~\ref{noconcmash}B shows that the predicted and observed~\cite{andrews} changes in $\con{NO}$ from its baseline value are within $\sim$10\% of each other.

\subsection{Model Predictions}
 
\begin{figure}[htbp]
\centering
\includegraphics[width=\textwidth]{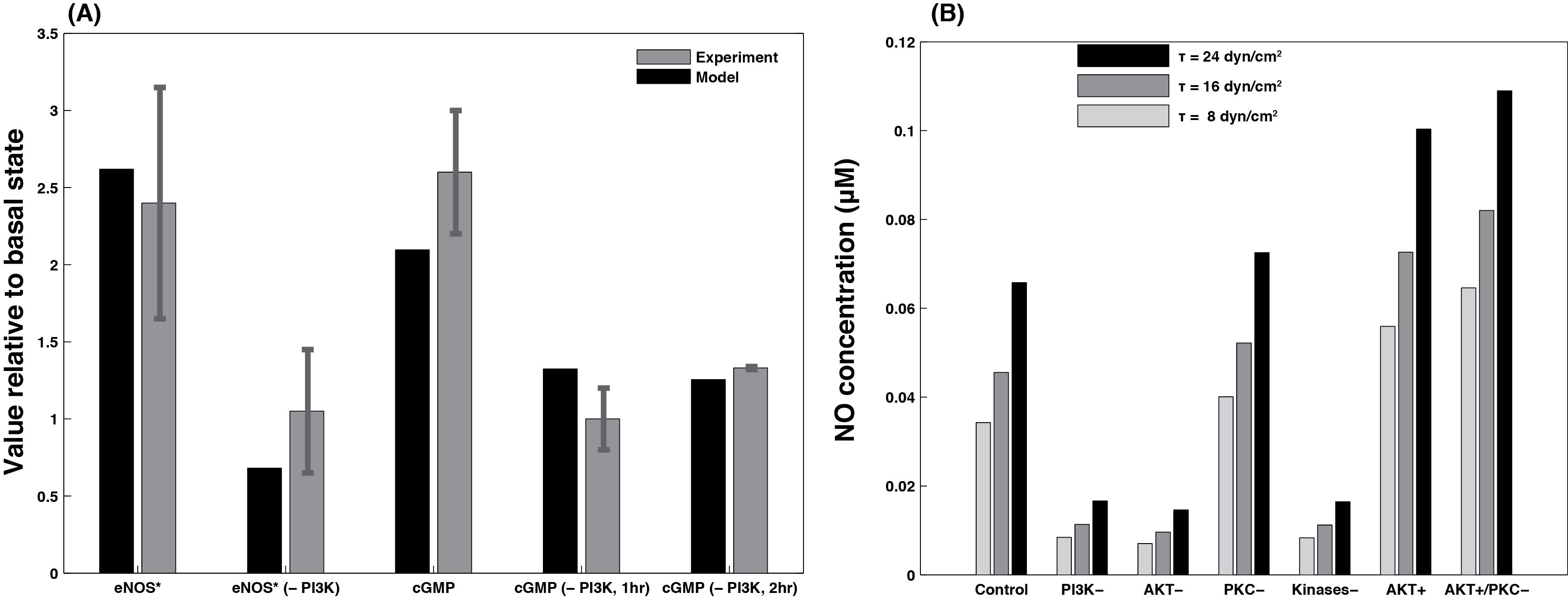}
\caption{\revise{\small Impact of modulation of protein kinase activity on NO production. \textbf{(A)} The predicted and observed eNOS phosphorylation by AKT, $\con{eNOS^*}$, at normal and completely inhibited kinase activity (PI3K and PI3K$^-$). Also shown is the corresponding effect on cGMP, with and without PI3K inhibition after 1  and 2 hours. The experimental data are from~\cite{dimmelernature}. \textbf{(B)} The predicted changes in steady-state $\con{NO}$, at three values of WSS, in response to elimination of PI3K activation (PI3K$^-$) and elimination of phosphorylation of eNOS by either AKT (AKT$^-$) or PKC (PKC$^-$). Also shown is the impact on steady-state $\con{NO}$ of the simultaneous elimination of eNOS phosphorylation by both AKT and PKC, as well as of the increase in AKT activity with (AKT$^+$) or without (AKT$^+$/PKC$^-$) PKC.}}\label{kinaseinhibition}
\end{figure}

\paragraph{Impact of inhibition of protein kinases on NO production.} The role of protein kinases in eNOS activation is typically studied by inhibiting either the kinases individually~\cite{matsubara, hirata} or PI3K, which results in the subsequent inhibition of both AKT and PKC~\cite{dimmeler, dimmelernature}. Figure~\ref{kinaseinhibition}A compares the model predictions with the experimental data from \cite{dimmelernature} for AKT-phosphorylated eNOS ($\con{eNOS^*}$, normalized with the concentration at $\tau = 0$) in response to a shear stress of $\tau = 12$ dynes/cm$^2$ applied for 1 hour (as stated in the methods in \cite{dimmelernature}; they also indicate that experiments were done at 15 dynes/cm$^2$ in the relevant figure legend, in either case, our results are in general agreement with their findings). Also shown are data from model vs experiment for the same, but with PI3K inhibited by wortmannin. Both model and experiment indicate that PI3K inhibition drastically reduces eNOS activation due to shear stress. Further, this inhibition of PI3K also impacts cGMP concentration; Figure~\ref{kinaseinhibition}B shows the predicted increase in cGMP concentrations over basal levels following 1hr exposure to a shear stress of 15 dynes/cm$^2$ is similar in both model and experiment. Further, the inhibition of PI3K by wortmanin induces reduction in cGMP concentrations to a similar degree in model and experiment. 

Figure~\ref{kinaseinhibition}B shows the impact of the complete inhibition of a) PI3K (PI3K-), b) AKT and c) PKC phosphorylation of eNOS (AKT- and PKC-) and d) both b) and c) occurring simultaneously (Kinases-), on predictions of steady-state NO concentration at WSS $\tau = 8, 16$ and $24$ dynes/cm$^2$. Inhibition of PI3K yields a significant ($\sim 70$ to 75\%) reduction in $\con{NO}$ over a range of $\tau$, while the inhibition of PKC alone results in a smaller but still significant ($\sim 10$ to 15\%) increase in $\con{NO}$. The predicted magnitude of reduction in $\con{NO}$ due to PI3K inhibition is in the vicinity of the data in~\cite{gallis} for reductions in NO synthesis: $\sim$ 70 to 75\% predicted versus 68\% reported. This finding is in agreement with other experimental studies, which found that  PI3K inhibition leads to decreased eNOS activity~\cite{dimmelernature, gallis} and that PKC inhibition elevates eNOS activity~\cite{hirata}. Blocking eNOS activation by AKT has a nearly identical effect as inhibiting PI3K; blocking PKC phosphorylation (and inactivation) of eNOS does little to retrieve this loss of NO production. These results indicate that AKT phosphorylation (and activation) of eNOS has a significantly larger net effect on endothelial NO production than does phosphorylation (and inactivation) by PKC. Finally, Fig.~\ref{kinaseinhibition}B indicates that increasing (doubling) AKT activity (by doubling the rate at which AKT phosphorylates eNOS, analogous to an AKT overexpressor model) significantly increases NO concentrations, an effect that is further enhanced by blocking PKC phosphorylation of eNOS. 

\subsection{Calcium-calmodulin dependence of eNOS activation and endothelial NO production}

\revise{Various stages of eNOS activation and NO production in endothelial cells are thought to be ``calcium-dependent'' or ``calcium-independent''~\cite{Balligand, dimmelernature,kuchanfrangos}. The sensitivity analysis reported below elucidates the dependency of NO production on calcium/CaM and AKT phosphorylation, the calcium-dependent and calcium-independent elements of the reaction cascade, respectively. In these simulations, the model was initialized using the default parameters specified in Table~\ref{table:kinetic_index}. Then, the concentrations of extracellular calcium and CaM were varied, while maintaining the WSS levels. This facilitates comparison of the model predictions and observational data from experiments, which consisted of simultaneous application of WSS and pharmacological modulation of the different pathways, such as calcium or AKT signaling.}

\begin{figure}[htbp]
\centering
\includegraphics[width=0.8\textwidth]{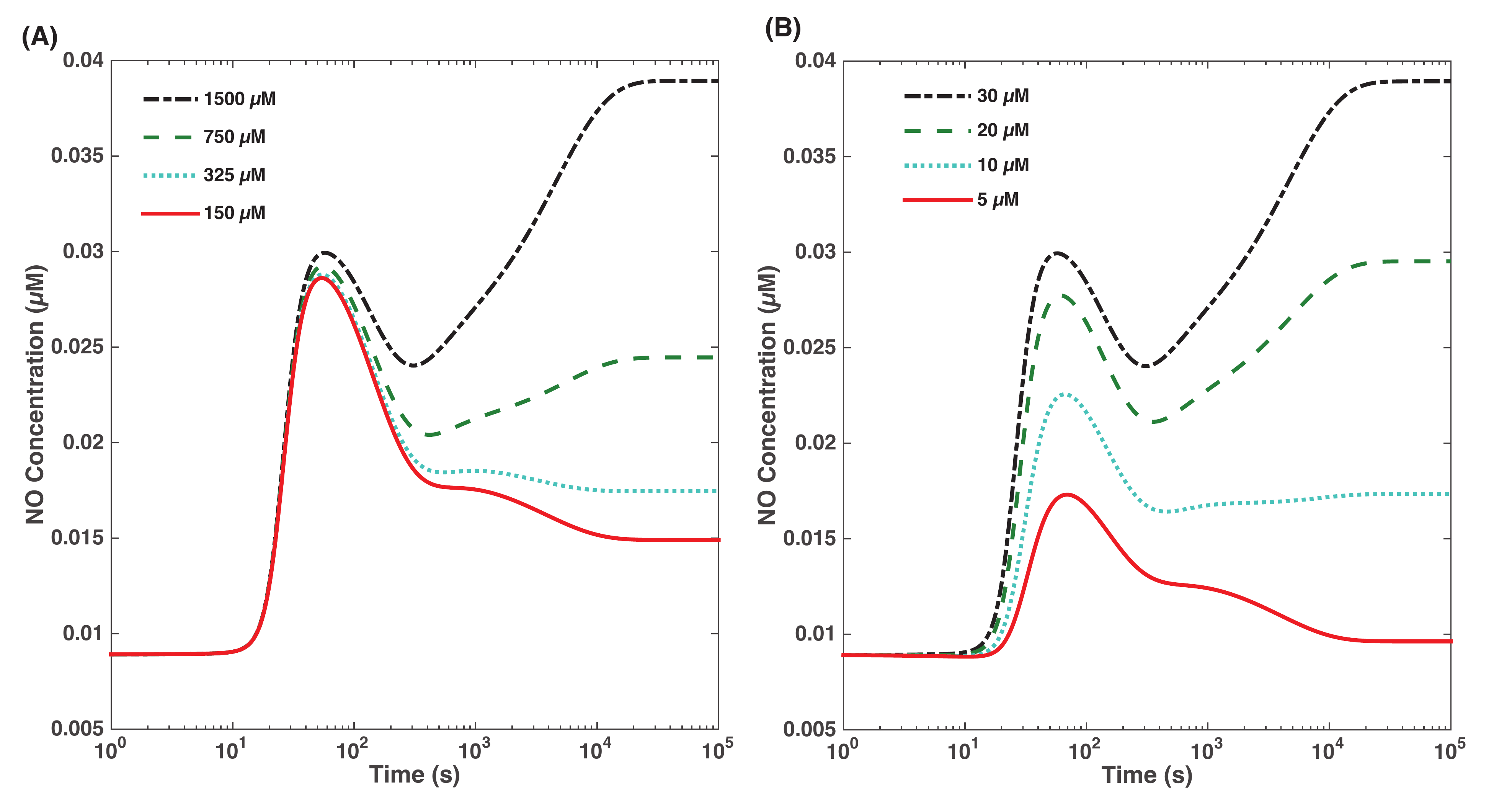}
\caption{(A) Effect of external calcium concentration on NO production. At high external calcium concentration, we observe a  biphasic dynamics of NO. When external calcium is depleted, the first phase of NO is unchanged (as this is largely driven by rapid release of calcium from internal stores) but the second phase of NO production is lost. (B) Effect of calmodulin concentration on NO production. When sufficient CaM is present, NO production displays biphasic kinetics. When CaM is depleted, a smaller first phase of NO is observed but the second phase is abolished. }\label{supextcalc}
\end{figure}

\revise{Figure \ref{supextcalc}A shows $\con{NO}$ and cytosolic calcium concentrations at WSS $\tau=12$~dynes/cm$^2$, for different concentrations of extracellular calcium. The decrease in the extracellular calcium concentration leads to the corresponding depletion of intracellular calcium, resulting in the reduction of endothelial NO production. The release of stored calcium causes an initial spike in NO production, which diminishes drastically at large times (at the time scale on which phosphorylation by AKT becomes an important driver of NO production), because Ca-CaM and thus eNOS-CaM concentrations fall as calcium is being depleted. Thus, while AKT activation/phosphorylation is itself independent of calcium \cite{dimmelernature, ozeki}, there is less substrate (eNOS bound to CaM) for phosphorylated AKT to act upon. Hence, phases of NO production that are apparently calcium-independent and driven mainly by AKT are in fact calcium-dependent. The reduction of NO production with depletion of extracellular calcium is in agreement with results in \cite{dimmelernature}, where chelation of extracellular calcium was shown to substantially reduce endothelial NO production, regardless of AKT activation. Similarly, when CaM concentration is decreased, both the first phase and second phase of NO production are affected but for very low CaM concentration ($\sim 5$~$\mu$M), the second phase of NO production is completely abolished. These results indicate that both early and later stages of NO production are calcium-dependent in different ways.} 

%

\section{Discussion and Conclusions}

We developed a biochemical model of the WSS-induced activation of eNOS in an endothelial cell (EC). The model includes three key mechanotransducers: MSICs, integrins and GPCRs. The reaction cascade consists of two interconnected parts. The first is rapid activation of calcium, which results in formation of calcium-calmodulin complexes, followed by recruitment of eNOS from caveolae. The second is phosphorylation of eNOS by PKC and AKT (both of which are activated by PI3K), which leads to further activation of eNOS. Our model also includes a negative feedback loop due to inhibition of calcium influx into the cell by cGMP. In this feedback, increased NO levels cause a rise in cGMP levels, so that cGMP inhibition of calcium influx can limit NO production.

The model was used to predict the dynamics of NO production by an EC subjected to a step increase of WSS from zero to a finite physiologically relevant value. To determine basal conditions, we ran the model at the steady-state regime with zero shear stimulation. At the basal conditions, $\con{IP_3} \approx 0$ and concentration of calcium stored in the smooth endoplasmic reticulum is at its maximal levels. Our model predicts that under these basal conditions there is a finite, non-zero level of NO production and finite, non-zero concentrations of AKT and PKC phosphorylated eNOS. These findings are in agreement with the observations~\cite{ozeki}. 

The model predicts a highly nonlinear, biphasic transient behavior of eNOS activation and NO production: a rapid initial activation due to the very rapid influx of calcium into the cytosol (occurring within 1 to 5 minutes) is followed by a sustained period of activation due to protein kinases, which are in turn activated by PI3K (Figs.~\ref{calciumcam}, and \ref{enoscamakt}). The predicted calcium- and kinase-dependent phases of eNOS activation are in agreement with the existing paradigm of the sequential steps of eNOS activation~\cite{Balligand, Shyy, mount}. Over large time periods, the enhanced activation of PI3K and subsequent activation (above basal levels) of AKT and PKC due to WSS are not major factors in eNOS activation. This is due to the apparent rapid decay of PI3K (and hence phospho- AKT and PKC) activity back to their basal levels. More experimental data are needed to clarify the behavior of PI3K following a cell's exposure to WSS and to fine-tune our model. 

To validate the model, we compared its predictions with both quantitative and qualitative experimental observations. \revise{The model reproduces the observed dependence of NO production rate (Fig.~\ref{newfigure4}) and NO concentrations (Fig.~\ref{noconcmash})} on wall shear stress, especially taking into account the significant amount of scatter in the available experimental data. The model reproduces the observed transient behavior of NO \revise{production rate} following a cell's exposure to shear stress (Fig.~\ref{newfigure4}B). The model predictions of both the increased levels of AKT phosphorylated eNOS and cGMP concentrations are within $\sim 10$ to 15\% of their measured counterparts (Fig.~\ref{kinaseinhibition}A).

Our model predicts that the inhibition of PI3K, which results in downregulation of both AKT and PKC, leads to a drastic reduction (over 70\%) in NO concentrations (see Figure~\ref{kinaseinhibition}); this is consistent with the observed behavior~\cite{gallis, dimmelernature}. This suggests that the stimulatory effect of AKT phosphorylation dominates the inhibitory role of PKC, supporting the perspective that AKT is by far the most important protein kinase in eNOS activation. This hypothesis is further supported by the model's prediction that inhibiting PKC alone results in a small but significant ($\sim 15$\%) increase in NO production; hence, PKC has a smaller net effect on eNOS activation than AKT. We are not aware of quantitative measurements of the extent of kinase inhibition as a function of eNOS activity, at different levels of shear stress. Yet our model does agree with the empirical evidence for strong up-regulation of eNOS activity by AKT and weaker, but still significant, down-regulation of eNOS by PKC.

To summarize, our model captures the following observed features of eNOS activation by WSS.

\begin{enumerate}
\item \revise{eNOS activation and NO production are \textit{always} dependent on calcium concentrations and the binding of eNOS with CaM. In other words, concentrations of calcium and CaM remain dependent variables for NO production/concentration and depleting or inhibiting either induces a precipitous decline in NO production. }

\item \revise{Certain elements of the reaction cascade that results in eNOS activation and NO production are reported to be calcium independent, as observed in ~\cite{dimmelernature,ozeki}. In our model, the coupling of AKT phosphorylation to calcium signaling is weak and, practically, calcium-independent. As a result, eNOS activation and NO production can be altered even if calcium signaling is unchanged. For example, inhibiting AKT phosphorylation while leaving calcium stimulation unchanged results in a sharp decrease in $\con{NO}$ (see Figs.~\ref{kinaseinhibition} and \ref{supextcalc}), in agreement with the experiment~\cite{dimmelernature}. }

\item \revise{Upon exposure to shear stress, once calcium reaches its steady state, NO production/concentration continues to change. Hence, there is a lag between calcium and NO dynamics; this is due to the relatively slow rate at which eNOS binds with CaM and is then phosphorylated by AKT. That does not imply independence of NO production from calcium; if calcium levels are changed during this slower phase of activation, NO production/concentrations will also change, albeit more gradually.}

\item Inactivation of AKT drastically reduces eNOS activity, whereas inhibition of PKC has a smaller, stimulatory effect on eNOS activation. While the elevation (above basal levels) of AKT activation due to WSS is not important over large time scales, the finite, basal activation of AKT is essential for maximal eNOS activation and NO production.

\item Concentrations of both NO and cGMP increase with WSS. Over a broad range of WSS this increase is highly non-linear, but within the physiologically relevant ranges of WSS (around {20~dynes/cm$^2$}) one can use a linear relation between [NO] and WSS~\cite{mashour, mcallister, sriramautoreg}.

\item The predicted steady-state and transient variations of NO production rates at different WSS levels are in general agreement with the observations, although the scatter in the reported data is large~\cite{andrews, kuchanfrangos, mcallister, kaur, mashour}. 

\item The predicted increase of $\con{cGMP}$ with $\con{NO}$ (Figure~\ref{kinaseinhibition}A) is consistent with the experimental data \cite{dimmelernature}. Hence the model supports a role of shear stress as a stimulator of vasodilation by quantifying the shear-induced NO production which, in turn, elevates a cGMP level, ultimately leading to vasodilation. 

\item Removal of WSS leads to an ultimate return to basal levels for all reactants, with the calcium transients occurring rapidly and the kinase-dependent transients following more slowly. The time scales over which the system returns to basal levels of [NO] are in a general agreement with the experiments~\cite{mashour}.

\end{enumerate}
\revise{Thus, sustained shear-induced endothelial NO production requires \textit{both} calcium signaling and AKT phosphorylation; the system can however be manipulated/modulated by inhibiting or promoting one pathway without changing the other.}

\revise{The presented model enhances the current understanding of the mechanistic and biochemical processes involved in the activation of eNOS and subsequent NO production in ECs. The model's predictions might be used to facilitate the design of experiments, which focus on inhibition of the reactants and mechanosensors involved in the NO production reaction cascade. This is relevant to several fields of biomedical research, e.g., cancer~\cite{weiming}, diabetes~\cite{creager} and heart disease~\cite{davies}, where the regulation of endothelial NO production has significant clinical applications.}

\paragraph{Model limitations.}
\revise{While our model captures many of the observed features of NO production in ECs, it has several  limitations.} It does not contain a mechanistic description of ECs. Instead, WSS acts as an input that triggers the reaction cascade resulting in eNOS activation. The lack of a mechanical model precludes the analysis of such factors as the role of viscoelasticity in determining the system's transient behavior. Consequently, our model cannot be used to reproduce experimental studies, which show that viscoelastic properties of the cytoskeleton have a profound effect on the mechanical behavior of an EC exposed to oscillatory or pulsatile shear. Future extensions of our model will combine it with a mechanical model of ECs. 

Our model accounts for three mechanosensors:  MSICs, GPCRs, and integrins. As experimental data for other mechanosensors (sodium and potassium ion channels, lipid rafts and vesicles, cytoskeletal remodeling, signaling via cadherins and other transmembrane proteins, etc.) become available, their effects can be incorporated into our model. Such enhancements of our model are facilitated by its modular structure.

\section*{Author Contributions}

KS performed research and wrote the paper. JGL led the revisions effort and edited the paper. PR helped with analysis and wrote the paper. DMT designed research and wrote the paper. All authors agree on the content of the paper.

\section*{Acknowledgements}

This work was supported in part by Defense Advanced Research Projects Agency under the EQUiPS program, the Air Force Office of Scientific Research under grant FA9550-12-1-0185 and by the National Science Foundation under grant DMS-1522799.

\bibliographystyle{plain}
\bibliography{enos}

\newpage
\setcounter{page}{1}
\renewcommand*{\thepage}{S\arabic{page}}

\setcounter{section}{0}
\renewcommand\thesection{S\arabic{section}}

\setcounter{equation}{0}
\renewcommand{\theequation}{S\arabic{equation}}

\setcounter{figure}{0}
\renewcommand{\thefigure}{S\arabic{figure}}

\setcounter{table}{0}
\renewcommand{\thetable}{S\arabic{table}}

\begin{center}
{\Large{\textbf{Supplemental Material for ``Shear-Induced Nitric Oxide Production by Endothelial Cells''} }}
\end{center}

\section{Model Parameterization}
\label{app1}

Most model parameters were taken directly from previous studies (see Table~\ref{table:kinetic_index}). The rest are estimated below.

\paragraph{GPCR activation and IP$_3$ production.} The variation of GPCR activation with WSS $\tau$ was fitted to the data reported in~\cite{chach}.  No IP$_3$ is produced at $\tau = 0$~\cite{plank} and, hence, there is negligible GPCR/G-protein activation. The data in~\cite{chach} suggest that G-protein activation reaches its maximum around $\tau = 15$ to $20 $~dynes/cm$^2$~\cite{chach} before plateauing. In Eq.~\ref{rstar}, the value of $\Lambda$, which determines shear stress at which G-protein activation reaches its peak, was selected as $\Lambda = 15 $~dynes/cm$^2$. Sensitivity to this parameter is discussed below. We set $K_{cp} = 0.002$~$\mu$M$^{-1}$, which at maximum G-protein stimulation results in a maximal level of IP$_3$ production = 0.00546~$\mu$M/s~\cite{plank}.

The total concentration of G proteins in the cell was calculated based on numbers from \cite{lemon, adams, riccobene}, implying $10^5$ G proteins in a cell volume of $5 \times 10^{-16}$~m$^3$ \cite{lemon}. This yields $\con{G} = 0.33$~$\mu$M in a cell. The parameter $\alpha$ in Eq.~\ref{ip3} was hence converted from the value used in \cite{lemon} for 10$^5$ G proteins to the value indicated in Table~\ref{table:kinetic_index} for $\con{G} = 0.33$~$\mu$M. The ATP-dependent parameter $M_\text{ATP}$ was estimated from \cite{plank} as $M_\text{ATP} = \phi_{\infty}/(\phi_{\infty}+k_c)$ = 0.7937 using the notations and reference values provided in~\cite{plank} for these quantities, under the assumption of a constant ``bulk" ATP concentration $\phi_{\infty}$. Lastly, the calcium dependence of IP$_3$ production was not considered, since previous studies (e.g., \cite{plank} and \cite{comerford}) parameterized this calcium dependence of IP$_3$ production such that terms canceled, resulting in an IP$_3$ production rate that was effectively independent of calcium. 

According to~\cite{lemon}, the term $r_r$ in Eq.~\ref{pip2}, which quantifies the extent of replenishment of PIP$_3$ by internal stores of phospholipids in the cell, can take values ranging from 10 s$^{-1}$ to 0.015 s$^{-1}$. Our predictions for $\con{NO}$ are relatively insensitive (a $< 1\%$ variation) to the value of this parameter.

\paragraph{CCE Inhibition.} For the low ($\mu$M)  concentrations of cGMP typically observed in ECs \cite{yang}, the data in~\cite{Kwan} suggest small, but measurable, decrease in CCE with increasing intracellular cGMP concentrations. At higher [cGMP], cGMP almost completely inhibits CCE; however we restrict our analysis to the range of [cGMP] that allows for small degrees of CCE inhibition by cGMP, with the rate of inhibition increasing linearly with [cGMP]. From the data in \cite{Kwan} for CCE inhibition by cGMP,  we estimate $\psi(\con{cGMP}) = 1 -  0.0075 \con{cGMP}$, i.e., $\sim$ 15\% reduction of CCE at $\con{cGMP} \sim 20$~$\mu$M.

\paragraph{PI3K activation.} On exposure to WSS, PI3K activity reaches its maximum of around 3.5 times its basal level~\cite{go, katsumi}. This gives  $a_\text{PI3K} = 2.5$ in Eq.~\ref{kpip2}. We explored a range of values for $\eta$, from 0.03~s$^{-1}$ (a very rapid decay of PI3K activity observed in~\cite{go}) to 0 (no decay). We also explored a range of values of $\delta$, which corresponds to the WSS at which maximal activation of PI3K occurs. Our model was found to be relatively insensitive to $\delta$; we set $\delta = 24$ dynes/cm$^2$ in all the  simulations. 

\paragraph{Ca$_4$-CaM binding.} We assumed the rate of Ca$_4$-CaM formation to equal the rate of calcium buffering by cytosolic proteins, as suggested in~\cite{jafri, plank}. This yields $k_\text{Ca$_4$-CaM} = 100$~s$^{-1}$, based on the value used in \cite{plank}. The maximum [Ca$_4$-CaM] at saturation levels of calcium was set to 4.5~nM/$\mu$M of the total available CaM~\cite{pereschini}. This gives a value of $\theta = 0.0045$ in Eq.~\ref{ca4cam}. Following~\cite{pereschini}, we set the Hill coefficient to $\beta = 2.7$.

\paragraph{eNOS-CaM binding.} The rate of dissociation of eNOS-CaM was set to $K_\text{CaM}^- = 0.01$~s$^{-1}$, based on the values reported in~\cite{mcmurry}. The maximum rate at which eNOS binds to CaM in the presence of caveolin was estimated by assuming that at basal conditions, 90\% of eNOS is in the inactive, non-CaM-bound state. This yields $K_\text{CaM}^+ = 7.5$~s$^{-1}$. The rate $K_\text{0.5CaM} = 3.0$~$\mu$M was estimated from the data in~\cite{michel} in the regime where [Cav] is significantly larger than [Ca-CaM] and, hence, [Ca$_4$CaM].

\paragraph{eNOS-protein kinase activation kinetics.} The rate of phosphorylation of eNOS by PKC was estimated from the data in~\cite{bredt}, under the assumption of first-order kinetics. This gives $k_\text{Thr}^+ = 0.02$~s$^{-1}$ (for full activation of all 100 nM of intracellular PKC) and $k_\text{Thr}^+/ k_\text{Thr}^- = 9.0$ in Eq.~\ref{eq:eNOScav}. The kinetics of eNOS binding with AKT, in presence of Hsp90, were estimated from the data in~\cite{takahashi} as $k_\text{eAKT}^\text{max} = 0.002$~s$^{-1}$ (for full activation of intracellular AKT) and $k_\text{eAKT}^\text{max} / k_\text{eAKT}^-  = 18.0$ in Eq.~\ref{eq:step2}, under the assumption of maximal activation of AKT in~\cite{takahashi}. Lower levels of AKT activation, i.e., $\con{AKT}/\con{AKT}_\text{tot} < 1$, shift this equilibrium to less AKT-phosphorylated eNOS. This estimate of the rate of AKT phosphorylation of eNOS results in an increase of $\con{eNOS^*}$ over its basal state following a step increase in WSS, which is comparable to that reported in~\cite{dimmeler} (see Fig.~\ref{kinaseinhibition}). An increase of similar magnitude (i.e., two- to three-fold) in eNOS phosphorylation was also measured in~\cite{kaur} following exposure to WSS.

\paragraph{NO production.} Based on a range of suggested values discussed in \cite{chenpopel}, we set $\con{eNOS}_\text{tot} = 0.04$~$\mu$M within the cell (see Table~\ref{table:kinetic_index}). Under the assumption of constant supply of L-Arg and oxygen, the rate of NO production, $R_\text{NO}$ in Eqs.~\ref{qno} and~\ref{rno}, depends solely on eNOS activity.  Hence, the ratio $k_\text{eNO} \con{O_2} / (k_\text{mNO}+\con{O_2})$ is assumed to be constant, $\Upsilon$ (see Eq.~\ref{o2const}).  We set $\Upsilon = 300$~s$^{-1}$ in Eq.~\ref{o2const}, which yields a basal rate of NO production $Q_\text{NO} \sim 3.4$~$\mu$M/s, a value consistent with the range of basal NO production rates in~\cite{chenpopel}. At physiological WSS values ($\tau \sim 20$~dynes/cm$^2$), NO production rates are found to be in the range of 20 $\mu$M/s, consistent with the  values suggested in previous studies~\cite{chenx, lamkin, chenpopel}. At WSS $\tau = 6$~dynes/cm$^2$, the predicted NO production rates are consistent (within 20\%) with the values suggested in~\cite{chenpopel}. These observations suggest that $\Upsilon = 300$~s$^{-1}$ yields the NO production rates that are consistent with those reported in the literature. Figures~\ref{newfigure4} and~\ref{noconcmash} demonstrate that the model's predictions of $\con{NO}$ are also consistent with their experimentally observed counterparts. The rate of NO scavenging was set to $\lambda_\text{NO} = 382$~s$^{-1}$, the rate at which RBCs adjacent to ECs scavenge NO \cite{sriramno, chenx}. 

\paragraph{Intracellular sGC concentrations.}  In Eq.~~\ref{eq:sgc}, we set $\con{sGC} = 0.1$~$\mu$M. This value was deduced from two observations: the total protein concentration in a cell is $\sim 0.2 \times 10^6$ proteins per $\mu$m$^3$~\cite{milo}, and the total amount of sGC protein in a variety of tissues is $\sim 400$~ng/mg~\cite{lewicki}. We verified that decreasing or increasing $\con{sGC}$ by an order of magnitude has a negligible effect ($< 1\%$) on $\con{NO}$ because the mass balance in Eq.~22 is dominated by the RBC scavenging term $\lambda_\text{NO} \con{NO}$.

\section{Supplemental Results}
\label{sec:sup-res}

Consider an endothelial cell at its basal state (WSS $\tau = 0$). We use our model to investigate the response of this EC to a step increase of WSS ($\tau = \tau_0 > 0$) applied at time $t = 0$. The amplitude and time scale of this response are controlled by a number of parameters, whose values are discussed in section~\ref{app1} and summarized in Table~\ref{table:kinetic_index}. Two of these parameters, $\delta$ and $\eta$ in Eq.~\ref{kpip2}, were estimated via a sensitivity analysis. This was done due to the lack of conclusive experimental evidence to facilitate parameter estimation for these quantities. 

\begin{figure}[htbp]
\centering
\includegraphics[scale=0.5]{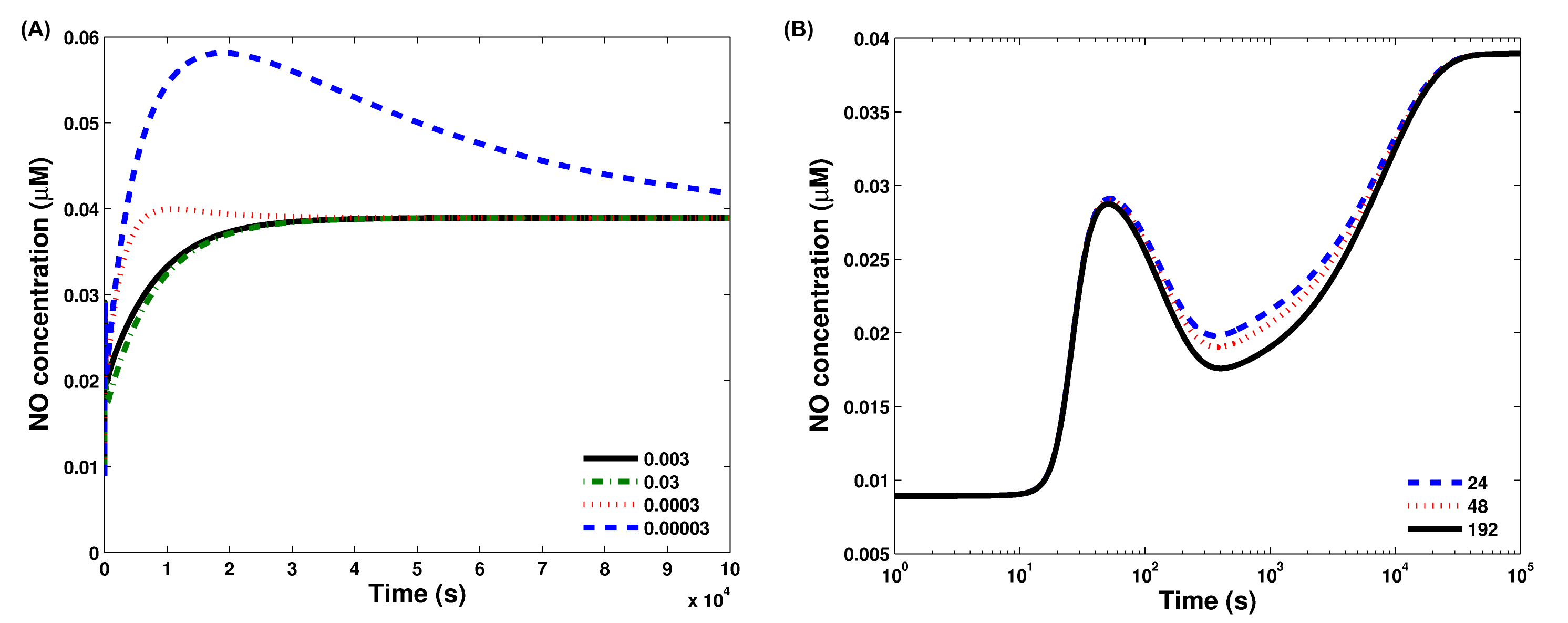}
\caption{Temporal variation of NO concentration produced by an endothelial cell in response to WSS $\tau = 12$~dynes/cm$^2$ for (a) $\delta = 24$~dynes/cm$^2$ and several values of $\eta$, and (b) $\eta = 0.003$~s$^{-1}$ and several values of $\delta$. These two parameters control the amplitude and time scale of the decay of PI3K activity to its basal level.}\label{etadelta}
\end{figure}

Experimental determination of their values is typically based on observations of the rate of decay of $\con{PI3K^*}$ to its basal state. Such estimates of the decay rate $\eta$ vary from $\eta \approx 0.003$~s$^{-1}$ (corresponding to a relatively fast time scale of about 5 min)~\cite{go} to $\eta \approx 0.0005$~s$^{-1}$ (corresponding to a larger time scale of about 30 min)~\cite{katsumi}. Our model enables one to discriminate between these estimates by propagating their effects through the reactive network. The outcome is presented in Figure~\ref{etadelta}, which shows the temporal variability of $\con{NO}$ in response to the applied WSS $\tau = 12$ dynes/cm$^2$ for several values of $\delta$ and $\eta$. Measurements of the shear-induced endothelial NO production~\cite{andrews, kuchanfrangos} reveal that $\con{NO}$ reaches its maximum value at the new equilibrium (steady state) corresponding to the applied WSS. This behavior is consistent with the $\con{NO}$ curves for $\eta \ge 0.003$~s$^{-1}$, while smaller values of $\eta$ (e.g., $\eta = 3.0 \times 10^{-4}$~s$^{-1}$ in Fig.~\ref{etadelta}) result in an unphysical intermediate maximum. In the remainder of this study we therefore use $\eta = 0.003$~s$^{-1}$, which yields the time scales for the decay of PI3K activity observed in~\cite{go}. For this value of $\eta$, the predicted dynamics of $\con{NO}$ is relatively insensitive to the choice of $\delta$ (Fig.~\ref{etadelta}b). We set its value to $\delta = 24$ dynes/cm$^2$, in the same range of magnitude as the WSS scaling parameter $\Lambda$ in Eq.~\ref{rstar}.

Relative insensitivity of our model to parameter $\delta$ suggests that the evolution of $\con{NO}$ over large time intervals after the EC's exposure to a step increase in WSS is largely unaffected by the shear-induced enhancement in activation of AKT and PKC. This is due to the rapid return of PI3K activity to its basal levels of excitation. At the same time, both AKT and PKC do play a crucial role in eNOS regulation as discussed below. Specifically, we will show that basal levels of AKT activation are both necessary and sufficient for a high level of eNOS phosphorylation and activation. 

\subsection{Parametric sensitivity of the dynamic behavior of NO concentrations}
\label{sec:sup-[NO]}

\begin{figure}[htbp]
\centering
\includegraphics[scale=1]{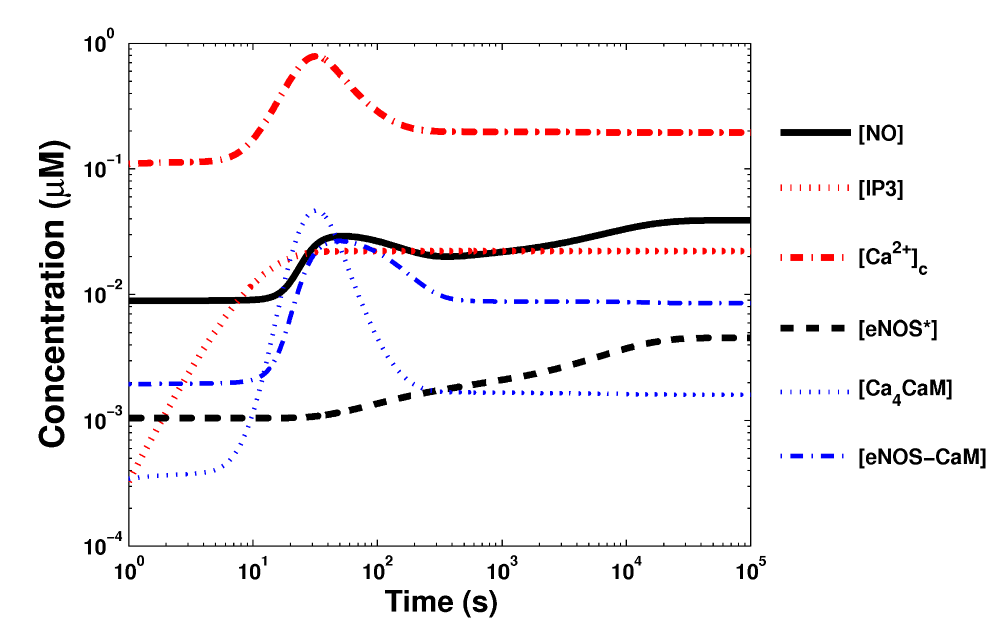}
\caption{\small Temporal variation of $\con{IP_3}$, $\con{Ca^{2+}}_\text{c}$, $\con{NO}$, $[\text{eNOS-CaM}]$ (concentration of eNOS bound to CaM, i.e., unphosphorylated by AKT), $\con{Ca_4CaM}$ and $\con{eNOS^*}$ (concentration of eNOS phosphorylated by AKT) for WSS $\tau = 12$~dynes/cm$^2$.}\label{alltransients}
\end{figure}

Figure~\ref{alltransients} identifies temporal scales of the variation of the components of the eNOS activation process by exhibiting the transient behavior of $\con{Ca^{2+}}_\text{c}$, $\con{Ca_4CaM}$, $[\text{eNOS-CaM}]$, $\con{eNOS^*}$ and $\con{NO}$ for WSS $\tau = 12$ dynes/cm$^2$. It elucidates the calcium-dependent and kinase-dependent phases of the eNOS activation cycle, with the calcium-dependent activation occurring very rapidly. The biphasic transient behavior is predicted to occur over a wide range of shear stress (see, e.g., Fig.~\ref{enoscamakt} in the main text) for eNOS activation and NO production. The two distinct phases of NO production correspond to the initial transient calcium response, and a more sustained but gradual increase in NO production following eNOS phosphorylation. The rapid calcium transient drives a spike in calcium-calmodulin complex formation $\con{Ca_4CaM}$. This results in enhanced recruitment of eNOS otherwise bound to caveolin, resulting in activation of the eNOS enzyme. During this early phase activation, eNOS phosphorylation by AKT plays a relatively minor role, as evidenced by the relatively slow increases in $\con{eNOS^*}$ (on the order of 10$^3$~s). The more gradual phase of the increased NO production/concentration starts when $\con{Ca^{2+}}_\text{c}$ reaches its steady state, which occurs on time scales on the order of 10$^2$~s. This steady-state value of $\con{Ca^{2+}}_\text{c}$ is elevated above its counterpart in basal, unstimulated cells. That, in turn, increases the steady-state amount of calcium-calmodulin complexes above their basal levels. Calmodulin-bound eNOS, eNOS-CaM, is then gradually phosphorylated by AKT, gradually increasing NO production as eNOS becomes phosphorylated and activated. 

\begin{figure}[htbp]
\centering
\includegraphics[scale=0.6]{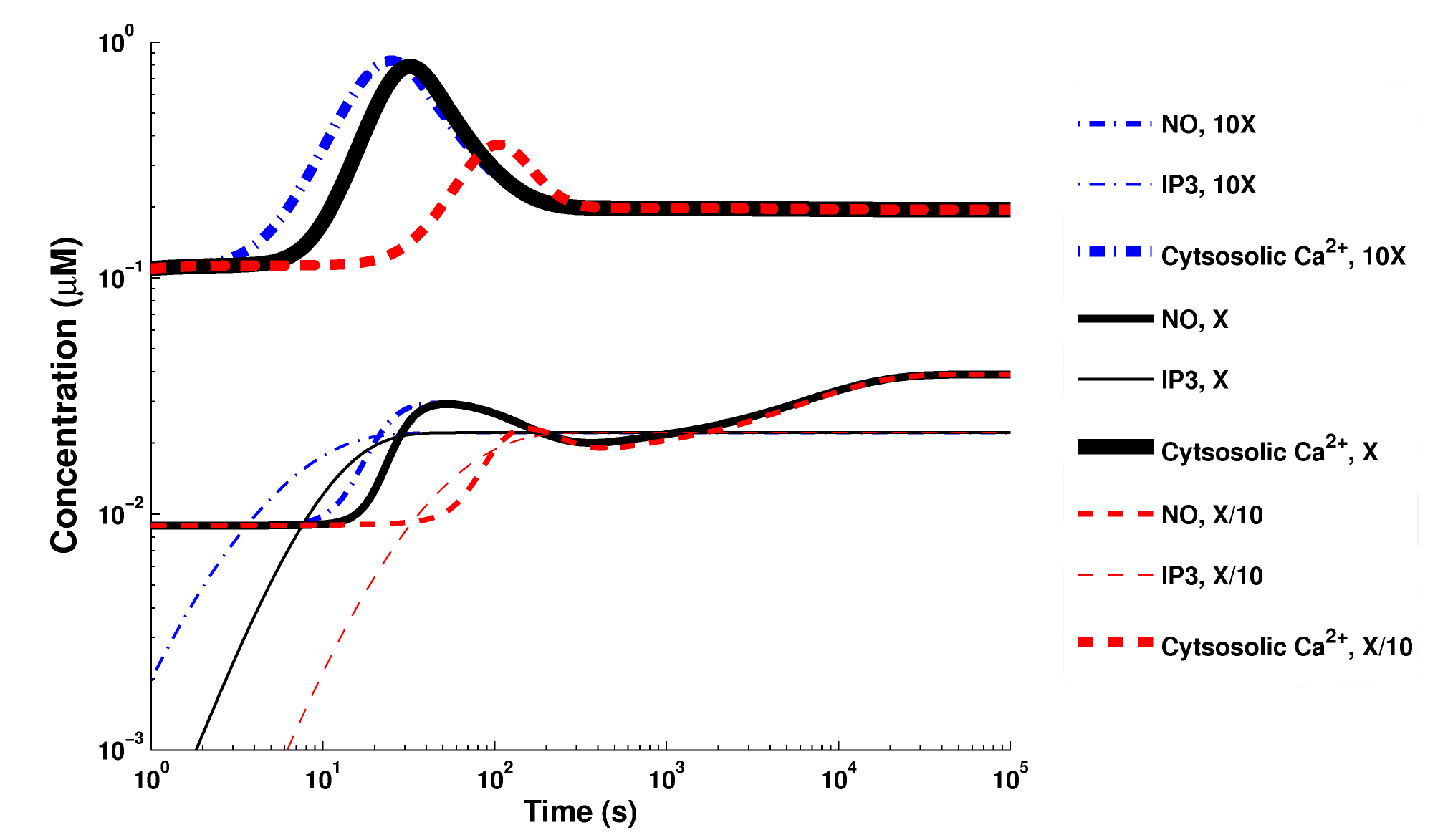}
\caption{\small Temporal variation of $\con{NO}$, $\con{IP_3}$ and $\con{Ca^{2+}}_\text{c}$, following a step-increase in WSS $\tau = 12$~dynes/cm$^2$, for several values of the IP$_3$ synthesis/degradation rates in Eq.~\ref{ip3}. 1X corresponds to the values of the rate constants reported in Table~\ref{table:kinetic_index}. X/10 and 10X corresponds to the values of these constants scaled by a factor of 1/10 and 10, respectively. }\label{ip3dynamics}
\end{figure}

To explore this dynamic behavior in greater detail, time scales of certain critical reaction processes were varied and their impact on the transient behavior of $\con{NO}$, following the exposure to WSS, was studied. The initial calcium transients are determined by the production of IP$_3$, which, in turn, mediates the release of stored calcium within the cell~\cite{plank}. The time scales of IP$_3$ synthesis and degradation are controlled by the forward ($r_f$) and backward ($\mu_1$) rates in Eq.~\ref{ip3}. Figure~\ref{ip3dynamics} exhibits the sensitivity of the biphasic behavior of NO concentrations to a \revise{tenfold increase and a corresponding decrease in $r_f$ and $\mu_1$ respectively, in response to a step-increase in WSS, such that the dynamics might change but not the steady-state response}.  The ten-fold reduction in $r_f$ modulates the early-time  increase in NO concentrations, because  a reduced rate of IP$_3$ synthesis decreases the magnitude of the early-time calcium transients.  The effects of the ten-fold increase in $r_f$ are less pronounced. At large times ($> 10^3$~s), variation in the IP$_3$ synthesis rate does not significantly impact NO concentrations, i.e., the steady-state cytosolic calcium concentration $\con{Ca^{2+}}_\text{c}$ and, therefore, reactants downstream of calcium, including the activated eNOS.

\begin{figure}[htbp]
\centering
\includegraphics[scale=0.8]{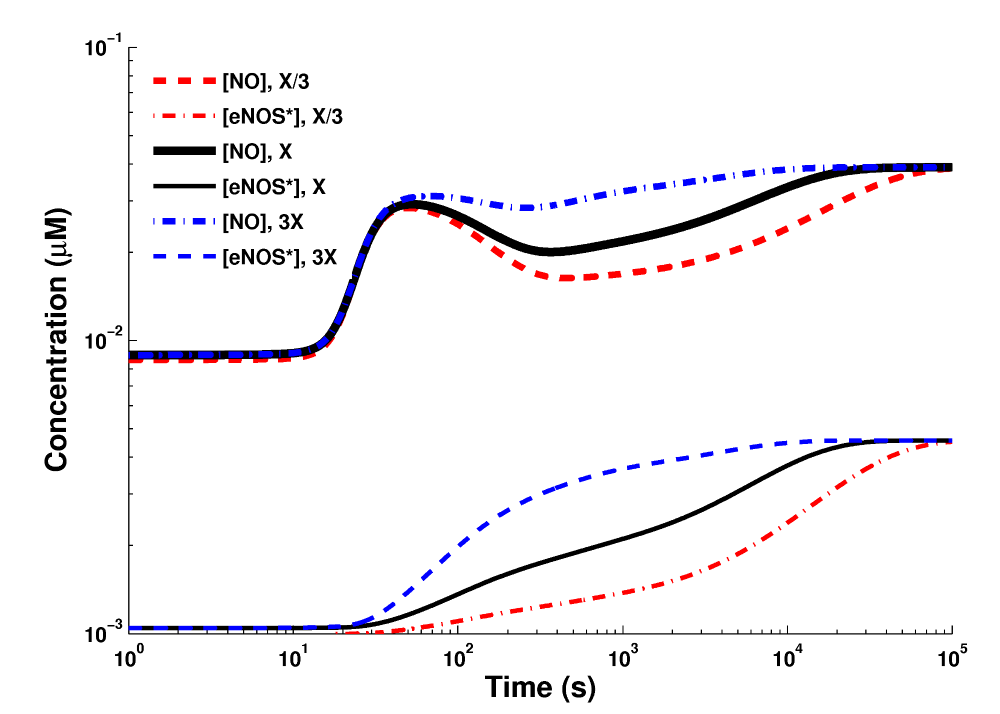}
\caption{\small  Temporal variation of $\con{NO}$ and $\con{eNOS^*}$, following a step-increase in WSS $\tau = 12$~dynes/cm$^2$, for several rates of AKT phosphorylation of eNOS (i.e., formation of [eNOS-CaM$^*$]) in Eq.~\ref{eq:step2}.  1X corresponds to the values of the rate constants reported in Table~\ref{table:kinetic_index}. X/3 and 3X corresponds to the values of these constants scaled by a factor of 1/3 and 3, respectively. }\label{aktdynamics}
\end{figure}

Figure~\ref{aktdynamics} exhibits the results of a similar sensitivity analysis for the dependence of transient NO concentrations on phosphorylation by AKT, described by Eq.~\ref{eq:step2}. The forward ($k^\text{max}_\text{AKT}$) and reverse ($k^-_\text{eAKT}$) rate constants for AKT phosphorylation of eNOS are scaled equally (modifying the rate at which eNOS is phosphorylated, but not the steady state, as both the forward and reverse rate constants are scaled equally). The rate at which phospho-eNOS is formed is effectively scaled, while other rates (such as Ca induction) are maintained constant. By construction, the steady-state amount of phosphorylated eNOS is not changed, but the rate at which the overall process proceeds has been increased. The faster the phosphorylation/dephosphorylation of eNOS occurs, the less of a ``hump" or biphasic behavior is seen, as the lag between the fast processes (IP$_3$ and Ca influx) and slow processes (AKT phosphorylation) has been reduced. The reverse occurs when the AKT phosphorylation is slowed down. The three-fold decrease in the rate of phosphorylation of eNOS amplifies the pronounced biphasic behavior of NO concentrations vs WSS, as the lag  between rapid calcium spike-driven NO production and the slower, sustained phase of AKT-eNOS phosphorylation-driven NO production widens. Conversely, the three-fold increase in the rate of eNOS phosphorylation by AKT dampens the biphasic behavior, due to reduction in this lag.

\begin{figure}[htbp]
\centering
\includegraphics[scale=0.8]{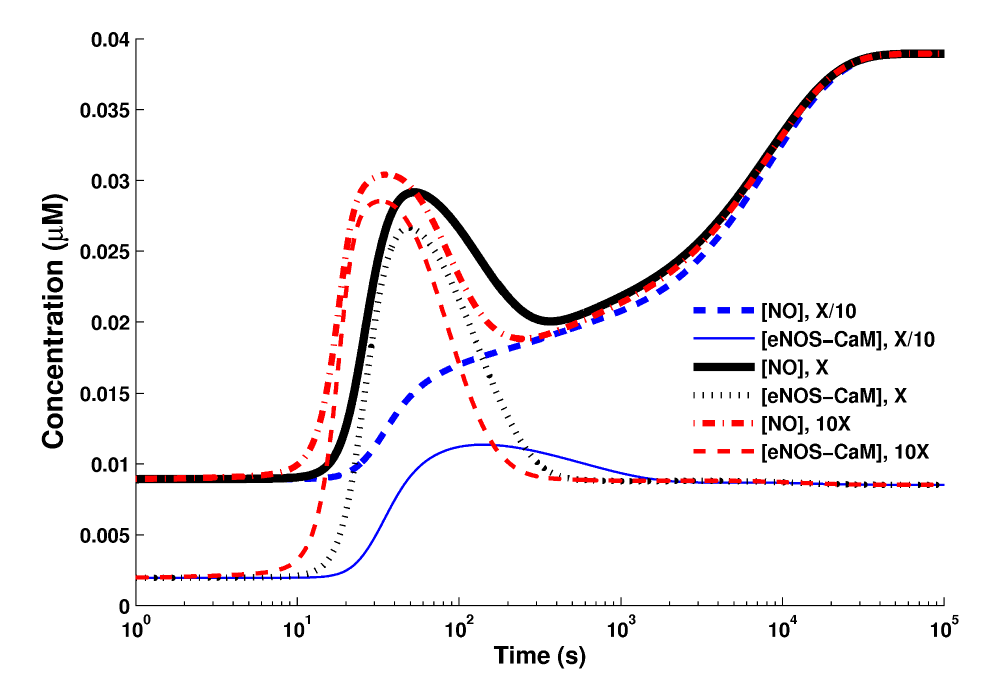}
\caption{\small Temporal variation of $\con{NO}$ and $[\text{eNOS-CaM}]$, following a step-increase in WSS $\tau = 12$ dynes/cm$^2$, for various of the forward ($k_\text{CaM}^+$) and backward ($k_\text{CaM}^-$) rates of formation of the eNOS-CaM complex (Eq.~\ref{eq:eNOS-Cam}). 1X corresponds to the values of the rate constants reported in Table~\ref{table:kinetic_index}. X/10 and 10X corresponds to the values of these constants scaled by a factor of 1/10 and 10, respectively.}\label{eNOSCaMdynamics}
\end{figure}

Figure~\ref{eNOSCaMdynamics} demonstrates the sensitivity of the predicted concentrations $\con{NO}$ and $[\text{eNOS-CaM}]$ to changes in the forward ($k_\text{CaM}^+$) and backward ($k_\text{CaM}^-$) rates of formation of the eNOS-CaM complex (see Eq.~\ref{eq:eNOS-Cam}). The ten-fold increase in the value of $k_\text{CaM}^+$  enhances (and accelerates) the biphasic response of NO concentrations to a step-increase in WSS, as the early spike in calcium (and thus a spike in concentration of calcium-CaM complexes) drives an even faster rate of recruitment of eNOS by CaM. The result is more activated eNOS and, thus, increased NO production. The opposite occurs when $k_\text{CaM}^+$  is reduced. As with IP$_3$, the impact on NO concentrations at larger times ($> 10^3$) is negligible.

\begin{figure}[htbp]
\centering
\includegraphics[scale=0.8]{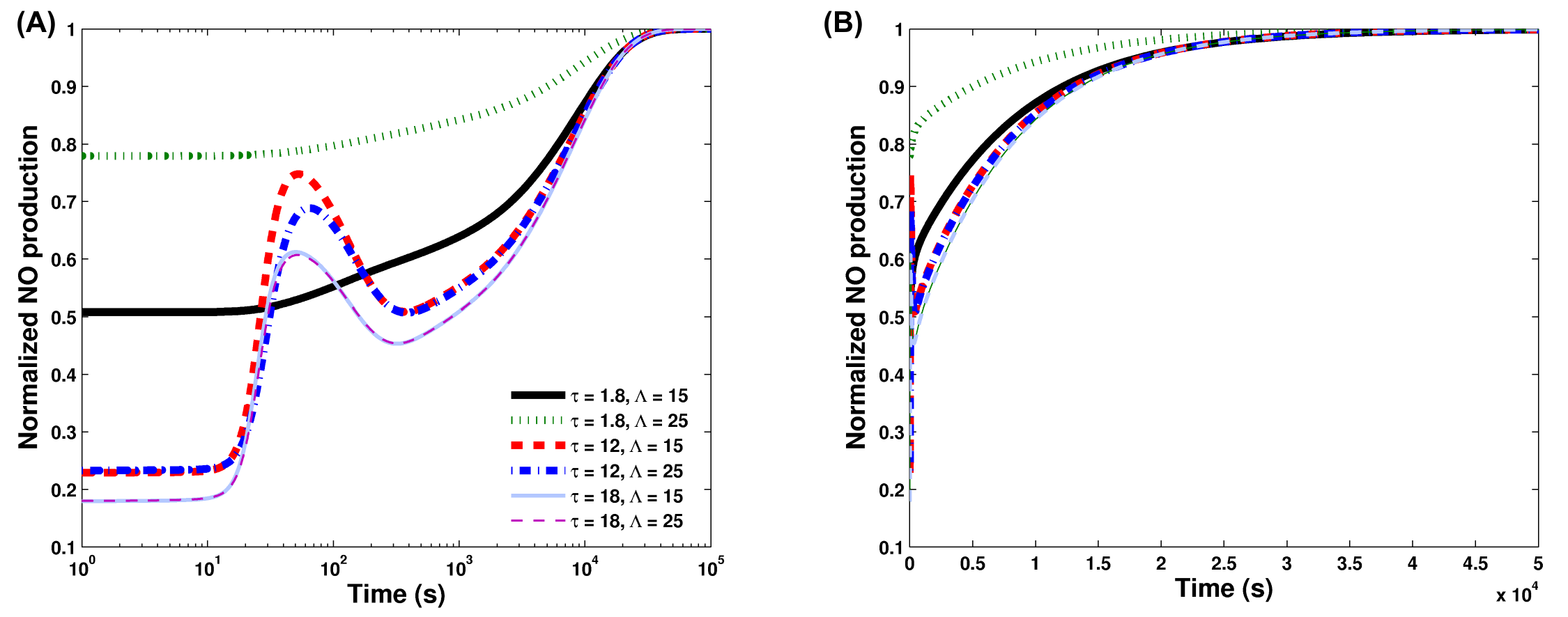}
\caption{\small Temporal variation of NO production rates, normalized with its steady-state value, at several values of WSS $\tau$ (in dynes/cm$^2$), for $\Lambda = 15$ and $25$~dynes/cm$^2$, plotted on both log (left) and linear (right) scales.}\label{taulambda}
\end{figure}

Figure~\ref{taulambda} shows the sensitivity of the model prediction of WSS-induced NO production to the parameter $\Lambda$ in Eq.~\ref{rstar}. For WSS $\tau \ge 12$~dynes/cm$^2$ (including physiologically relevant levels of $\tau \sim 20$~dynes/cm$^2$), while at lower values of WSS (e.g., a basal level $\tau =1.8$~dynes/cm$^2$) the choice of $\Lambda$ does affect the NO production. The value of $\Lambda=15$~dynes/cm$^2$ reported in Table 1 is at the low end of the physiologically relevant range, and predicts an $\approx 50\%$ increase in NO production from its basal level at $\tau =1.8$ dynes/cm$^2$. This prediction is close to experiment~\cite{mcallister}, where an increase in WSS $\tau$ from 0 to 1.8 dynes/cm$^2$ results in a nearly 60\% increase in NO production at steady state.

\subsection{Impact of feedback due to cGMP}

\begin{figure}[htbp]
\centering
\includegraphics[scale=0.5]{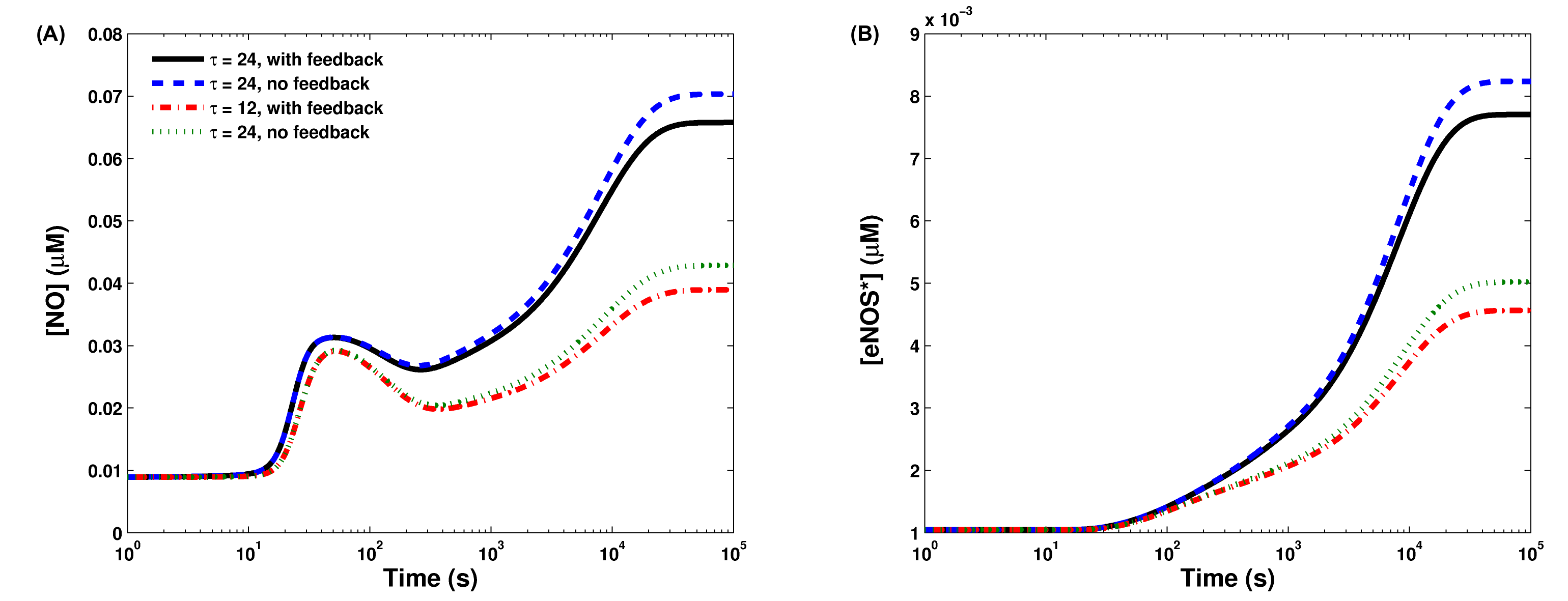}
\caption{\small Temporal variation of $\con{NO}$ and $\con{eNOS^*}$, with and without the cGMP feedback, at WSS $\tau = 12$ and 24 dynes/cm$^2$.}\label{feedback}
\end{figure}

According to Eqs.~\ref{cce} and~\ref{psi}, as well as the experimental evidence~\cite{Kwan}, the capacitative calcium entry (CCE) decreases as $\con{cGMP}$ increases. At large concentrations of cGMP ($\sim 1$~mM), the CCE is completely inhibited and, consequently, cytosolic calcium is strongly restricted. The physiologically relevant concentrations of cGMP predicted by our model, $\con{cGMP} \sim 10$ to $20$~$\mu$M, correspond to a much smaller degree of calcium inhibition. This raises a question about the significance of the feedback role of cGMP in the endothelial NO production at these cGMP concentrations. Figure \ref{feedback} shows $\con{NO}$ and $\con{eNOS^*}$, with and without feedback, at two levels of WSS $\tau = 12$ and $24$~dynes/cm$^2$. The cGMP feedback has a relatively minor ($\approx 5\%$ at the physiological levels of WSS, which corresponds approximately with $\tau = 24$~dynes/cm$^2$) effect on steady and transient NO concentrations and on eNOS activation levels. This suggests that the cGMP feedback  contributes little to regulation of endothelial NO production and that the negative feedback may even be neglected if the minor ($\sim 5\%$) errors this potentially introduces are not considered significant.

\subsection{Temporal variation of PIP$_3$ concentration}

\begin{figure}[htbp]
\centering
\includegraphics[scale=0.8]{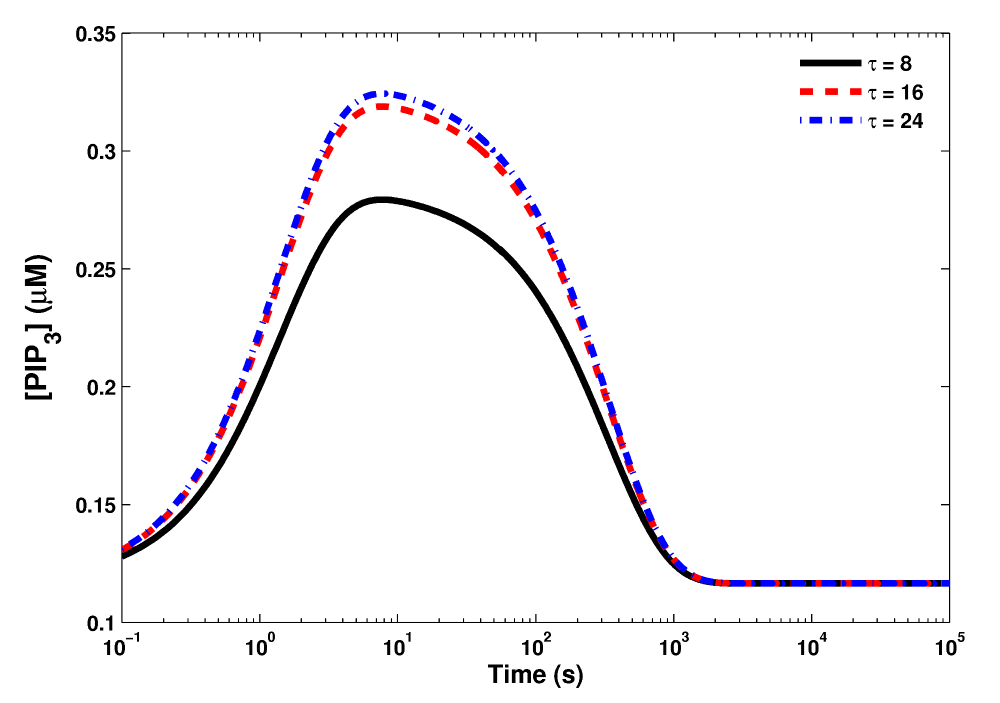}
\caption{\small Temporal variation of $\con{PIP_3}$ at different values of WSS $\tau$ (in dynes/cm$^2$). }\label{pip3new}
\end{figure}

At short-to-intermediate times (up to $10^3$~s), $\con{PIP_3}$ exhibits significant (three-fold or larger) growth (Fig.~\ref{pip3new}). Given the dependence of the reaction rate constants $k_\text{AKT}^+$ and $k_\text{PKC}^+$ on $\con{PIP_3}$ (see Eq.~\ref{eq:rates}), this renders an analytical solution of the AKT and PKC activation equations (see Eq.~\ref{eq:15}) unfeasible.

\subsection{Basal (initial) conditions}

The basal (resting) condition for an endothelial cell is defined by steady-state concentrations of all the reactants at WSS $\tau = 0$. The initial concentrations of reactants were thus computed by solving the steady-state version of Eqs.~\ref{gpcr}--\ref{eq:last} with $\tau = 0$. This gives (in units of $\mu$M) $\con{IP_3} = 0.0$, $\con{Ca^{2+}}_\text{c} = 0.10$, $\con{Ca^{2+}}_\text{s} = 2824$, $\con{PIP_2} = 10$, $\con{Ca_4CaM} = 0.00026$, $\con{eNOS_\text{cav}} = 0.029$, $\con{eNOS_\text{cav}^0} = 0.0078$, $\con{PIP_3} = 0.116$, $[\text{eNOS-CaM}] = 0.0019$, $\con{eNOS^*} = 0.001$, $\con{cGMP} = 3.8624$, $\con{NO} = 0.0089$. 

The basal value of $\con{IP_3} = 0.0$ $\mu$M reflects the lack of activation of the GPCR-PIP$_2$-IP$_3$ cycle in the absence of mechanical stimulus. In this steady state, the the smooth endoplasmic reticulum (ER) maintains a large reserve of stored calcium (with a basal value of $\con{Ca^{2+}}_\text{s}$). For the basal state to be viable, a small perturbation of the system from this state must dissipate in a finite amount of time; in other words, the stored calcium concentration  $\con{Ca^{2+}}_\text{s}$ must return to its basal value. This condition is enforced by the presence of ``leak current" $k_\text{leak} \con{Ca^{2+}}_\text{s}^2$ in Eq.~\ref{calcium}, which governs the calcium balance within the EC. Several studies, e.g.,~\cite{comerford, david, plank07}, neglected this leak current, which is present in the original calcium balance equation~\cite{plank, wiesner}. This omission results in an unstable initial condition to which the system never returns, even after the removal of stimulus. Hence, we argue that the leak current for stored calcium must be present in the equations for calcium dynamics.

Our calculated initial conditions imply that a little over 90\% of eNOS is initially bound to caveolin (with $\sim$20\% phosphorylated at Thr-495 by PKC). This is consistent with the proportion of inactive eNOS reported in~\cite{ju}. Though small, there exist detectable levels of both NO production and AKT phosphorylated eNOS at basal conditions. This finding is in agreement with the experiments~\cite{ozeki} that found a measurable degree of eNOS activity at basal conditions i.e., in the absence of WSS.

\section{Methods}

The system of equations presented in this study was analyzed as a coupled system of ordinary differential equations (ODEs) in MATLAB. This system was solved using the \texttt{ODE15s} MATLAB solver, optimized to handle stiffness that stems from the presence of a wide range of time scales. The solver's relative tolerance was set to $10^{-9}$ or lower. The code was checked for numerical/logical errors by verifying whether quantities supposed to be conserved (such as total amount of eNOS in all forms) were in fact conserved at all time points. The MATLAB code utilized is included as a separate supplement.

When experimental values were not explicitly stated in the text, tables, or figures of the cited articles, the data were estimated from the corresponding data plots. Estimation of graphically presented data was done using the Getdata Graph Digitizer software. These experimental data reported in Fig.~\ref{newfigure4} are reported in Tables~\ref{table:exp4a_1}--\ref{my-label}.

\begin{table}[!!h]
\centering
\caption{Data from \cite{kuchanfrangos}, where the raw data was presented in (nmol/mg protein/hr). These data are shown as squares in Fig.~\ref{newfigure4}A. }
\label{table:exp4a_1}
\begin{tabular}{|l|l|l|}
\hline
Shear stress & Raw value & Normalized value \\ \hline
0 & 0.7 & 1.00 \\ \hline
1.8 & 1.7 & 2.43 \\ \hline
6 & 2.2 & 3.14 \\ \hline
12 & 2.7 & 3.86 \\ \hline
25 & 5.2 & 7.43 \\ \hline
\end{tabular}
\end{table}

\begin{table}[!!h]
\centering
\caption{Data from \cite{kaur}, where the raw data was presented in (pmoles/10$^5$ cells/hr). These data are shown as circles in Fig.~\ref{newfigure4}A.}
\label{table:exp_4a_2}
\begin{tabular}{|l|l|l|}
\hline
Shear stress & Raw value & Normalized value \\ \hline
0.00 & 732.00 & 1.00 \\ \hline
4.00 & 1957.00 & 2.67 \\ \hline
\end{tabular}
\end{table}

\begin{table}[!!h]
\centering
\caption{Data from \cite{kanai}, where the raw data was presented in (pmol/s). These data are shown as triangles in Fig.~\ref{newfigure4}A. }
\label{table:exp4a_3}
\begin{tabular}{|l|l|l|}
\hline
Shear stress & Raw value & Normalized value \\ \hline
0.20 &	0.45	& 1.00 \\ \hline
1.00	& 0.48 & 1.07 \\ \hline
3.00 &	0.86	& 1.91 \\ \hline
5.00 &	1.31 &	2.91 \\ \hline
10.00 &	2.44 &	5.42 \\ \hline
\end{tabular}
\end{table}
\begin{table}[!!h]
\centering
\caption{Experimental data from \cite{tsao} used to generate Fig.~\ref{newfigure4}B. We used reduced Values (subtracting ``baseline" NO$_x$, since there is some non-zero NO present at zero shear, presumably due to some nitrate or nitrite already in culture media; the lowest measurement of NO$_x$ was at 2 hrs with no shear, which was defined as the baseline)}
\label{my-label}
\begin{tabular}{|c|c|c|c|c|c|c|}
\hline
Time (s) & NO$_x$      & No$_x$              & NO$_x$ (- baseline) & No$_x$ (- baseline)  & No$_x$ normalized & No$_x$, normalized     \\ \hline
         & No shear & Shear  & No shear            & Shear & No shear       & Shear \\ \hline
      & & 12 dynes/cm$^2$ & & 12 dynes/cm$^2$ & & 12 dynes/cm$^2$   \\ \hline
6264.00  & 1.74     & 3.10                & 0.57                & 1.93                &0.04 &	0.14\\ \hline
4212.00  & 1.17     & 2.79                & 0                   & 1.62                & 0.00&	0.11\\ \hline
6444.00  & 1.79     & 3.61                & 0.62                & 2.44                &0.04 &	0.17         \\ \hline
10980.00 & 3.05     & 5.29                & 1.88                & 4.12                & 0.13 &	0.29        \\ \hline
12672.00 & 3.52     & 8.85                & 2.35                & 7.68                & 0.17 &	0.54         \\ \hline
20520.00 & 5.70     & 15.31               & 4.53                & 14.14               & 0.32	& 1.00      \\ \hline
\end{tabular}
\end{table}

\section{Additional Discussion}

\paragraph{Model limitations.}

Following the standard practice in modeling biochemistry of ECs~\cite{plank, plank07, comerford, wiesner, wiesner1, david}, we assumed that all the reactions occur in a well mixed cytosolic medium, i.e., ignored the effects of spatial heterogeneity.  The homogeneity assumption is justified since the diffusion time scales of proteins within a cell's cytosol are usually fractions of a second~\cite{phelps}, whereas even the rapid calcium transients considered in our model occur at the time scale of 100~s. 


Our model does not account for protein kinases other than AKT and PKC. While the latter are generally regarded as the most important kinases that phosphorylate eNOS, other kinases such as PKA and MAPK also interact with eNOS~\cite{Balligand, mount}. 
As more information about the extent to which PKA impacts eNOS activity becomes available, this kinases can be added to our model to account for, e.g., up-regulation of eNOS due to phosphorylation at Ser-633.

Our model does not include protein phosphatases, such as PP2A and calcineurin. The model assumes a constant rate of dephosphorylation of eNOS with respect to both AKT and PKC, since there is relatively little information about how WSS modulates phosphatase activity. Changing levels of PP2A and calcineurin is expected to alter the rates of dephosphorylation of eNOS.
The reaction cascade presented in our model can be further extended by incorporating vasoactive molecules, such as bradykinin and insulin, to  represent the activation of receptors by these signaling molecules.   

Finally, our model ignores possible effects of WSS on transcription of various proteins central to the reactions involved, especially mRNA for eNOS, AKT, PKC and CaM. The modeling assumption of a constant supply of necessary substrates and proteins implies that RNA transcription and protein production respond to keep total protein concentrations approximately constant. 
Future work will expand the system of reactions to account for activation of various transcription factors. Among other elements, this would necessitate incorporating a model of WSS-induced cAMP production and PKA activation, followed by activation of transcription factors such as CREB, leading to calculations for the rate of production of important proteins that participate in the eNOS reaction cascade. 

\paragraph{Need for experimental evidence.}

While we conclude that the broad conceptual framework of how eNOS is activated by WSS is largely complete, various elements of this framework are still not wholly understood. Areas where more experimental data are needed include:
\begin{enumerate}
\item The extent to which inhibition of AKT and PKC impacts NO production and eNOS activation at different levels of WSS. Dose response curves relating the extent of kinase inhibition to both eNOS activation and NO production would help to clarify the kinetics of eNOS phosphorylation.

\item The relative roles of Ca$_3$CaM and Ca$_4$CaM in the recruitment of eNOS into an eNOS-CaM complex. We assumed that Ca$_4$CaM plays the primary role in recruitment of eNOS into eNOS-CaM complex, however more experimental data are needed to clarify this issue.

\item The extent to which PI3K and FAK are activated by mechanical stimulation, in both transient and at steady-state regimes. There is at present a paucity of data on this topic, particularly with regards to the transient behavior of PI3K activation.

\item The inhibitory effect of cGMP on calcium influx. We are aware of only one study~\cite{Kwan} that records the variation of CCE rates over a broad range of cGMP concentrations. The data presented suggest that a negative feedback provided by cGMP is likely unimportant at normal physiological conditions in ECs. The lack of such feedback would greatly simplify the task of modeling NO transport and coupling this model to upscaled transport models in order to simulate vascular mechanics~\cite{sriramautoreg}.

\item The impact of varying cytosolic ATP concentrations on the rates of phosphorylation of eNOS by AKT and PKC. This is especially a concern since WSS is known to increase endothelial ATP levels~\cite{barakatatp}; however, very little data are available on the relationship between WSS-induced ATP production and eNOS activation.  

\item The role and kinetics of various protein phosphatases in the dephosphorylation of eNOS. Very little information is available about the extent to which WSS impacts activity of phosphatases, especially PP2A and calcineurin. 
\end{enumerate}

\end{document}